\documentclass[
  aps,
  prl,
  reprint,
  twocolumn,
  superscriptaddress,
  floatfix,
  nofootinbib,
]{revtex4-1}

\usepackage{amsmath,amssymb,amsthm,mathtools}
\usepackage{bm}
\usepackage{bbm}
\usepackage{dsfont}
\usepackage{braket}
\usepackage{physics}

\usepackage{graphicx}
\usepackage[caption=false]{subfig} 

\usepackage[english]{babel}
\usepackage{microtype}            
\usepackage[normalem]{ulem}       
\usepackage{comment}

\usepackage[dvipsnames]{xcolor}
\usepackage[
  bookmarks=true,
  colorlinks,
  linkcolor=OrangeRed,
  urlcolor=NavyBlue,
  citecolor=RoyalBlue
]{hyperref}
\hypersetup{breaklinks=true,hidelinks,hypertexnames=false}      
\definecolor{mygold}{rgb}{0.93,0.59,0.13}
\definecolor{mypurple}{rgb}{0.49,0.18,0.56}

\definecolor{mygreen}{rgb}{0.25,0.5,0.25}

\begin{document}

\author{Volker Karle}
\email{vkarle@issp.u-tokyo.ac.jp}
\affiliation{Institute of Science and Technology Austria, Am Campus 1, 3400 Klosterneuburg, Austria}
\affiliation{Institute for Solid State Physics, University of Tokyo, Kashiwa 277-8581, Japan}

\author{Oriana K. Diessel}
\affiliation{Department of Physics, Harvard University, 17 Oxford Street Cambridge, MA 02138, USA}
\affiliation{ITAMP, Harvard-Smithsonian Center for Astrophysics, Cambridge, MA 02138, USA}

\author{Vasil~Rokaj}
\affiliation{Department of Physics, Villanova University,
             Villanova, Pennsylvania 19085, USA \looseness=-1}
\affiliation{ITAMP, Harvard-Smithsonian Center for Astrophysics, Cambridge, MA 02138, USA}

\author{Ceren B.~Da\u{g}}
\email{ceren_dag@g.harvard.edu}
\affiliation{Department of Physics, Indiana University, Bloomington, Indiana 47405, USA}
\affiliation{Department of Physics, Harvard University, 17 Oxford Street Cambridge, MA 02138, USA}
\affiliation{ITAMP, Harvard-Smithsonian Center for Astrophysics, Cambridge, MA 02138, USA}

\title{Hybrid light-matter boundaries of graphene in a chiral cavity}

\begin{abstract}
We establish 
the bulk-boundary correspondence for graphene in a circularly polarized cavity. Using exact diagonalization, transfer-matrix analysis, and introducing the Dirac-Jaynes-Cummings model, we identify chiral edge modes in the light-matter hybridization gaps, some bright with finite photon number. Particularly, analytical expressions are derived for the localization length and mode velocity of bright edge modes. Density matrix renormalization group on hybrid electron-photon tensors reveals that these gaps survive at finite electron density, leading to the first non-perturbative study of graphene in a chiral cavity, and connecting single- and many-electron pictures. We find that the time evolution converts dark edge states into bright unidirectional currents, which can provide an experimental signature of hybrid band topology, in addition to the many-body charge gaps.
\end{abstract}

\maketitle
\textit{Introduction.}~Controlling topological phases with light has become a defining ambition of material science over the last decades~\cite{RevModPhys.93.041002,Sentef2018,DeGiovannini2022APR}.
Periodic driving can induce
Berry curvature and non-zero Chern numbers on otherwise trivial bands~\cite{OkaAoki2009,Lindner2011,mciver2020light,Wintersperger_2020,Topp2021}, but the required high‐intensity illumination inevitably heats the lattice and limits coherence times~\cite{Oka}. Cavity quantum electrodynamics (QED) offers a conceptually different route: enhanced vacuum fluctuations of a confined mode reshape the electronic structure in \textit{equilibrium}, without an external drive~\cite{Frisk_Kockum_2019,hubener2021engineering,Schlawin2022,garcia2021manipulating,Bloch2022}.
Experiments have demonstrated artificial gauge fields for exciton-polaritons in microcavity lattices~\cite{Karzig2015,Schine2016} and ultrastrong light-matter coupling of cavity photons to Landau level excitations in a two-dimensional electron gas has been observed, leading to the formation of Landau polaritons~\cite{ li2018vacuum, scalari2012ultrastrong, RokajPRX, LangeRecordPolaritons} and modifying transport \cite{paravicini2019magneto, bartolo2018vacuum,doi:10.1126/science.abl5818, CiutiHopping, rokaj2023topological,enkner2024enhancedfractional,xue2025cavitymediatednonlinearlandau,PhysRevB.106.205114,andolina2025quantumelectrodynamicsgraphenelandau,yang2025quantumhalleffectchiral}. Complementarily, vacuum‐induced quantum anomalous Hall (QAH) gaps have been predicted for graphene, transition-metal dichalcogenides, and moiré heterostructures~\cite{PhysRevB.84.195413,PhysRevB.99.235156,PhysRevB.107.195104,nguyen2023electronphoton,PhysRevLett.132.166901,Dag2024,PhysRevB.109.195173,yang2025emergent,ghorashi2025tunabletopologicalphasesmultilayer} with novel chiral cavity designs to achieve time-reversal symmetry breaking vacuum \cite{suarez2024chiral,PhysRevB.109.L161302,Tay2024,kulkarni2025realizationchiralphotoniccrystalcavity}.

Yet the defining hallmark of topological matter, \textit{bulk–boundary correspondence}, is unverified in cavity quantum materials. Previous studies either treated light classically \cite{PhysRevA.91.043625,PhysRevB.90.115423} or focused on bulk gaps and Chern numbers \cite{PhysRevB.99.235156,Dag2024,ghorashi2025tunabletopologicalphasesmultilayer}, leaving uncharted whether the cavity photons hybridizing with Dirac electrons can generate topologically robust hybrid edge modes. In graphene, a chiral cavity promotes each Bloch band into an infinite ladder of photon replicas, bounded from below due to the existence of a stable vacuum. In such a hybrid band structure, light-matter interactions give rise to avoided crossings in the Brillouin zone (B.Z.), i.e.,~gaps that carry a non-zero Berry phase \cite{Dag2024}. Here we find that the light-matter interaction induced gaps support chiral edge modes of hybrid character, and these gaps survive at finite electron density.

For a 
topological mass, one may anticipate a dark electronic edge mode connecting the two valleys. 
The resonant photon-exchange gaps studied here are, however, distinct: they occur where electronic Dirac states of adjacent photon sectors hybridize, so including the photonic Hilbert space is essential. The corresponding topological boundary modes found in this work are therefore light-matter hybrid edge states with finite cavity occupation.

\begin{figure*}[!t]
\centering
\includegraphics[width=2.0\columnwidth]{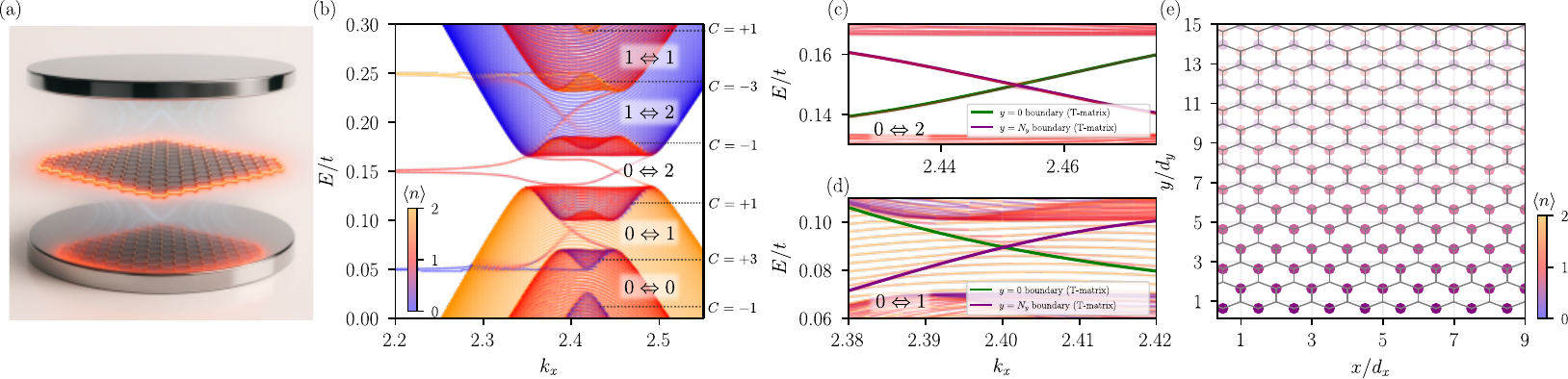}
\caption{\textbf{Graphene strip in a chiral cavity with its bright and topological edges.} (a) Conceptual set‐up: a graphene ribbon sits at the electric‐field antinode of a right‐handed cavity. The circularly polarized vacuum field breaks time‐reversal symmetry, imprinting a unidirectional edge channel (red glow) along each ribbon boundary.
(b) Single-particle spectrum for a zigzag ribbon of $N_y=1600$ sites, cavity frequency $\omega_c = 0.1t$, and coupling $g = 0.025t$ (Hilbert space truncated to two photons for convenience). The color scale shows the photon number expectation value of the state $\langle a^\dagger a\rangle$. Hybridization with the $n=0,1,2$ photon replicas opens interaction gaps, which host at least two chiral edge modes. Those involving a net photon exchange become strongly entangled (“bright”) and thus carry a finite number of photons.
(c–d) Edge dispersions from the semi‑infinite T‑matrix formalism, zoomed into the two‑photon (c, upper) and one‑photon (d, lower) exchange gaps. Colored lines overlay the finite‑ribbon data from (b), demonstrating that the analytic T‑matrix reproduces the same bulk–boundary correspondence, see ~\cite{supp}.
(e) Real‐space edge‐state profile at $k_x=2.4$: the transparency of the surface plot gives $|\psi(x,y)|^2$, while the color scale maps the local photon occupation. At the boundary, the state is in an equal superposition of $n=0$ and $n=1$ photons, yielding $\langle n\rangle\approx0.5$.}
\label{fig:fig1}
\end{figure*}

We develop a semi-analytic $T$-matrix (TM) method~\cite{wielian2025transfer,PhysRevB.93.134304,PhysRevB.48.11851} that proves the presence of these edge states in a semi-infinite lattice. We then introduce an analytical model that captures the nontrivial topology in light-matter hybridization points 
that was only numerically detected before~\cite{Dag2024,ghorashi2025tunabletopologicalphasesmultilayer}: 
the Dirac-Jaynes-Cummings (DJC) model extended from JC model in quantum optics~\cite{shore1993}. Using the continuum version of DJC model, we derive closed-form expressions for the edge current velocity, localization length and photon occupation. 
Furthermore, we reveal that turning on the cavity coupling in real time converts a dark electronic edge state into a bright hybrid edge state with finite photon number that propagates unidirectionally, which may be used as an experimental signature of cavity-induced topology. {Finally, we perform density matrix renormalization group (DMRG) where the photon mode is not integrated out, opposite to perturbative approaches~\cite{PhysRevB.99.235156,Dag2024,ghorashi2025tunabletopologicalphasesmultilayer,Rubiographene2025} showing a charge gap opening at finite density and supporting the relevance of our single-electron results to a many-body system.}

\textit{Graphene on a zigzag strip.}~The graphene Hamiltonian on a honeycomb lattice is $H(\bm{r}) = t \sum_{\bm{r},\bm{n}} \left(c^{\dagger}_{\bm{r}} c_{\bm{r+\delta_n}} + \text{h.c.}\right)$ with the bond vectors $\boldsymbol{\delta}_1 = \frac{\alpha}{2}\left(\sqrt{3}, 1\right), \boldsymbol{\delta}_2 = \frac{\alpha}{2}\left(-\sqrt{3}, 1\right),  \boldsymbol{\delta}_3 =\alpha \left(0,-1\right)$ where $\alpha$ is the lattice distance~\cite{bernevig2013topological} and $c_{\bm r}$ are fermionic operators at position $\bm r$ following $\lbrace c^{\dagger}_{\bm r},c_{\bm r'} \rbrace = \delta_{\bm r \bm r'}$. Following the recent advances in chiral cavity design based on InSb magnetoplasma which can suppress one of the polarizations \cite{Tay2024,kulkarni2025realizationchiralphotoniccrystalcavity}, we work with a quantized vector potential describing a single-polarization cavity field $\hat a\equiv \hat {a}_R$, where $[ \hat a,\hat a^{\dagger}] = 1$, and $\mathbf{e}\equiv \mathbf{e}_R = (1,\textrm{i})/\sqrt{2}$, ${\mathbf{A}}=\sqrt{\frac{1}{2\epsilon_0\mathcal{V}\omega_c}}\left[\mathbf{e}^* {a}^{\dagger}+\mathbf{e} {a}\right]$, where $\mathcal{V}=\chi \left(2\pi c/ \omega_{c}\right)^3 $ \cite{paravicini2019magneto} is the effective cavity volume with confinement parameter $\chi$ and single-mode cavity frequency $\omega_c$. For graphene, the tight-binding Hamiltonian from the minimal coupling agrees with the Peierls substitution expanded to the first order in the dimensionless light-matter interaction strength $g$. Hence we use the Peierls substitution for graphene on a zigzag strip, Fig.~\ref{fig:fig1}(a), and derive the Hamiltonian~\cite{supp} [$\hbar=1$]
\begin{eqnarray}
 \mathcal{H}&=& H_{\rm 0} -\frac{\textrm{i} t g}{\sqrt{2}}{\sum_k\sum_{j=1}^{N_y/2}} \biggl[ f_3(\hat a) c_{2j,k}^{\dagger}c_{2j+1,k} +  c_{2j,k}^{\dagger}c_{2j-1,k}  \notag \\
 &\times &\left(f_1(\hat a) e^{\textrm{i} k \sqrt{3}/2}+
f_2(\hat a) e^{-\textrm{i} k \sqrt{3} /2}\right)\biggr] + \rm h.c.,\label{eq:circ_real}
\end{eqnarray}
with
\begin{eqnarray}
H_{\rm 0} &=& \omega \left(\hat a^{\dagger}\hat a + \frac{1}{2} \right) + t{\sum_k\sum_{j=1}^{N_y/2}} \biggl[{c_{2j,k}^{\dagger}c_{2j+1,k}}\label{eq:circ_real_b} \\
&+&c_{2j,k}^{\dagger}c_{2j-1,k} \left(e^{i k \alpha \sqrt{3}/2}+
e^{-i k \alpha \sqrt{3} /2}\right) + \rm h.c.\biggr].\notag
\end{eqnarray}
Here, $c_{j,k}$ are fermionic annihilation operators  at position $j$ along the strip with momentum $k$ (in the units of $\alpha^{-1}$) in the orthogonal direction obeying $\lbrace c^{\dagger}_{j,k},c_{s,k'} \rbrace = \delta_{kk'}\delta_{js}$, and photonic function at honeycomb bond $\bm{\delta}_l$, $f_l(\hat a) = \left[ (\hat a+\hat a^{\dagger})\delta_{l,x} + \textrm{i}(\hat a-\hat a^{\dagger})\delta_{l,y}\right]$. $N_y$ denotes the number of atomic rows, so $j=1,\ldots,N_y/2$ labels
unit cells in Eqs.~\eqref{eq:circ_real}-\eqref{eq:circ_real_b}.
We distinguish two different regimes in Eq.~\eqref{eq:circ_real}. (i) The exact diagonalization (ED), the TM analysis and DJC theory solve the single-electron problem where hybridization across the photon copies is treated exactly. This is the setting in which we establish the topological boundary modes. (ii) At finite electron density, the photon mode is coupled to all electrons, mediating electronic interactions. The DMRG results in Fig.~\ref{fig:fig2} show that the Dirac and photon-exchange gaps underlying the single-particle boundary modes lead to robust electronic charge gaps in the many-body solution of Eq.~\eqref{eq:circ_real}.

\textit{Bright topological edge modes in the single-particle picture.}~ 
ED of ribbons up to $N_y=1600$, and TM analysis for a semi-infinite system shows that the edge modes survive at the boundary.  
For this, we extended the TM framework of Ref.~\cite{PhysRevB.93.134304} to our hybrid model with local Hilbert space dimension $|\mathcal H_{\mathrm{loc}}|=2(n+1)$, where $n$ is the maximum allowed photon number in numerics, and solve the Schrödinger equation in the $T$-matrix form, see ~\cite{supp}, which gives the semi-infinite solutions including the edge modes.

In Fig.~\ref{fig:fig1}(b) we show both the edge modes and the dispersive bulk bands with ED: the 
coupling to a chiral cavity splits the Dirac cone into a ladder of hybrid bands. The latter were defined to be a coherent superposition of electron and photon degrees of freedom with topological light-matter hybridization gaps in the electronic B.Z., hence labeling each hybrid band with a Chern number \cite{Dag2024}. We observe that the photon-exchange gaps
contain chiral edge states. These edge states are the most visible connecting two valleys, however they actually originate from the aforementioned topological light-matter hybridizations at a valley, which are hidden by the dispersive bulk bands in Fig.~\ref{fig:fig1}(b) [see for example energies $E/t \sim 0.1$]. Since finite ribbons resolve momenta only down to $k\!\sim\!2\pi/ N_y$, the edge modes residing at some energies, e.g.,~$E/t \sim 0.1$, seem to be strongly hybridizing with the dispersive bulk bands. These edge modes, however, are robust, as we systematically prove: The first signature of their robustness comes from their existence in the semi-infinite system. The resulting green and purple curves of the TM method in Fig.~\ref{fig:fig1}(c,d), lie on top of the ED data within line width, proving that the set of chiral edge modes is a result of the topological bulk in the single-particle picture, not a finite-size artifact, and persist throughout the photon ladder. Furthermore, these edge states exhibit an exponentially localized spatial profile in $y-$direction as seen in Fig.~\ref{fig:fig1}(e). The exponential decay of the probability is also plotted in Fig.~\ref{fig:fig4}(d) for different light-matter interaction strength $g$, and the localization length $\xi$ in $\vert \psi \vert^2 \propto e^{- y/\xi}$ is analytically derived in the next sections.

Colorbars in Fig.~\ref{fig:fig1} denote the average number of photons in the states. We call the edge states with finite photon number \textit{bright edge modes}.  
The lowest energy pair of edge modes that traverses the B.Z. between two valleys at $E/t\sim 0.05$ and crosses in between is vacuum-like, which we call \text{dark edge modes}. The presence of dark edge modes is evidence of bulk-boundary correspondence in the chiral vacuum induced Chern insulator with Chern number $C=1$ \cite{PhysRevB.84.195413,PhysRevB.99.235156,Dag2024}. Let us also note that bulk-boundary correspondence holds in the higher energies too: The Chern numbers of the bands from vacuum to higher energies for a photon number cutoff of $n=2$ are $C=-1,3,1,-1,-3,1$ \cite{Dag2024}. The first set of bright edge modes at energy $E/t \sim 0.1$ lies between the second and third bands. The
Chern number of the bands below these chiral edge modes is then $C_{1-2}=2$. Consequently, we count a pair of bright edge modes in each valley in Fig.~\ref{fig:fig1}(b). The gap at $E/t \sim 0.15$ hosts three crossings, with one at each valley and one in the middle of valleys, and the Chern number of the bands below sums up to $C_{1-3}=3$.

\begin{figure}[!tbp]
\centering
\includegraphics[width=1.00\columnwidth]{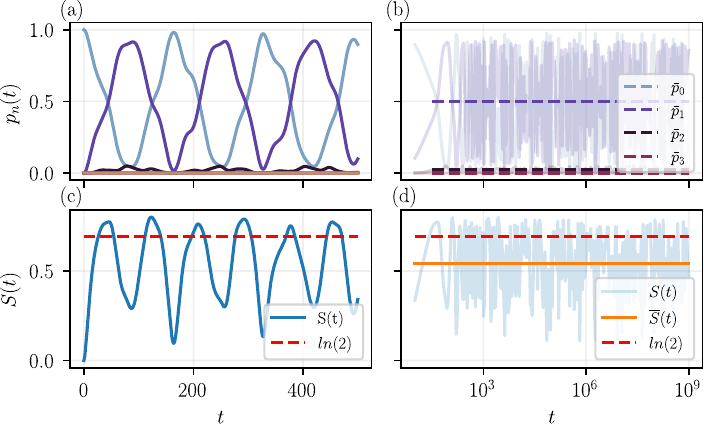}
\caption{\textbf{Coherent formation of a bright chiral edge mode from a dark edge state and build-up of light–matter entanglement.}
Time evolution is launched from a \textit{dark} edge state, vacuum Fock state, at fixed momentum $k_x = 2.4$ (same system parameters as in Fig.~\ref{fig:fig1}).
(a,b) Photon number distribution $p_n(t)$ in relatively short and long times. The initially product state quickly spreads over the neighboring Fock sectors as the bright chiral edge mode is populated.
(c,d) von Neumann entropy $S(t)$ of the reduced photonic density matrix in relatively short and long times. $S(t)$ oscillates around the singlet value $\ln 2$
(red dashed line), signaling a maximally entangled cat state of 1 photon and 1 electron. The orange line is the long-time average of $S(t)$ in (d).}
\label{fig:fig3}
\end{figure}

\textit{Coherent quench dynamics of bright edge modes.}~To investigate the single-particle edge-mode dynamics, we initialize the system in a \textit{dark} edge eigenstate at \(t=0\) with
\(\ket{\psi_0}\) of \(H_0\) at the nodal momentum \(k_{x,0}\) and suddenly
switch on $g$. This mimics, for instance, a setup where a movable cavity mirror is brought closer to the other mirror. The state
\(\ket{\psi(t)} = e^{-iHt}\ket{\psi_0}\) is propagated with a Krylov
time evolution scheme~\cite{rackauckas2017differentialequations}.  Tracing out the electronic sector gives the reduced
photonic density matrix
\(\rho_{\mathrm{ph}}(t)=\Tr_{\mathrm{e}}\dyad{\psi(t)}\).
Its diagonal elements $p_n(t)=\mel{n}{\rho_{\mathrm{ph}}(t)}{n}$ for $n \neq 0$ start being populated, where $\sum_{n=0}^{\infty}p_n(t)=1$ holds, signaling the dynamical
formation of a \textit{bright} edge state in Fig.~\ref{fig:fig3}(a). We observe coherent Rabi-like oscillations between vacuum and the 1-photon Fock state in this edge mode. Simultaneously the von-Neumann
entropy $S(t)=-\Tr\!\bigl[\rho_{\mathrm{ph}}(t)\ln\rho_{\mathrm{ph}}(t)\bigr]$ rises from \(0\) to \(\ln2\) after a single-photon exchange and oscillates
around $\ln 2$ with a frequency set by
$\Omega\simeq2tg$ for the parameters of Fig.~\ref{fig:fig3}(c). $S(t)$ is maximal when $p_{n=0}(t)\sim p_{n=1}(t)\sim 0.5$ (a cat-like state), so the oscillations of $p_n(t)$ show up in  $S(t)$.
The oscillations remain phase coherent over the simulated window because the initial state overlaps with only a few bright edge eigenstates.

\begin{figure}[!tbp]
\centering
\includegraphics[width=1.00\columnwidth]{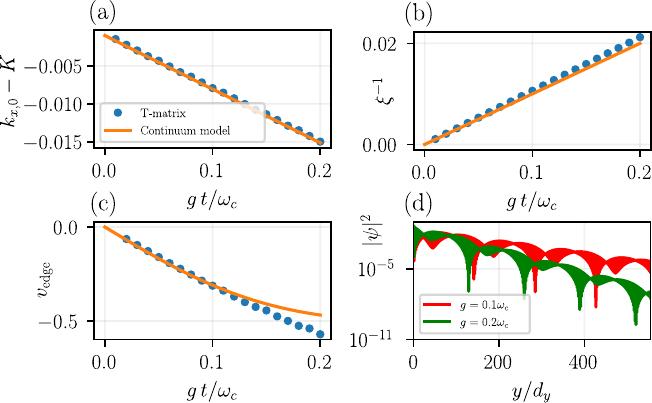}
\caption{\textbf{Localization length and the velocity of a bright edge mode.} (a) For $\omega_c=0.1 t$, we compare $k_{x,0}$, the momentum of the crossing between left and right localized edge mode, calculated using both the TM method for a semi-infinite graphene ribbon and the continuum DJC model. Similarly, in (b) the
inverse localization length $\xi^{-1}$ at $k_{x,0}$ and in (c) the edge mode velocity determined by these two methods. The localization length decreases with increasing light-matter interaction strength $g$, while the magnitude of the velocity increases.
(d) The bright edge mode probability decays as $|\psi|^2\propto e^{-2y/\xi}$.}
\label{fig:fig4}
\end{figure}

\textit{DJC Model and the continuum edge theory.}~
Linearizing the lattice model near one valley yields the Dirac–Jaynes–Cummings Hamiltonian in the bulk
\(H_{\mathrm{DJC}}(\mathbf k)=v_F(k_x\sigma_x+k_y\sigma_y)
        +\omega_c a^{\dagger}a
        +g_D(\sigma_+a+\sigma_-a^{\dagger})\),
where $v_F=\tfrac{3}{2}\,t$ is the Fermi velocity for the bond convention above (with $\alpha=1$) and $g_D$ defines the light–matter interaction scale
\(g_D=\sqrt2\,v_F g=\tfrac{3}{\sqrt2} t\,g\).
Although the DJC model is in general not analytically solvable, we can use a simple truncation to capture the physics of the gap openings due to light-matter hybridizations, as detailed in the SM. For the gap associated with 1 photon exchange, we truncate the Hilbert space to the 4-dimensional fixed-momentum subspace spanned by photon states $|n\rangle$ and $|n+1\rangle$, and by electronic states $\vert\downarrow\rangle_\mathbf{k}$ and $|\uparrow\rangle_\mathbf{k}$ in the sublattice basis, \(\Psi_{\mathbf k}=(c_{A,\mathbf k},c_{B,\mathbf k})^{T}\). Projecting the DJC Hamiltonian on this subspace leads to a $4 \times 4$ matrix which can be diagonalized and predicts that at momentum $k_\text{1p} = \omega_c/(2v_F)$ away from the $\mathbf{K},\mathbf{K'}$ points, a gap opens with size $\frac{\Delta_{\rm 1p}}{\omega_c}(\gamma)=\sqrt{2}\,\sqrt{1+\gamma-\sqrt{\,1+\gamma+\gamma^2\,}}$
where $\gamma=(g_D/\omega_c)^2(1+n)$. For $\gamma \ll 1$, i.e., $g_D\sqrt{n+1}\ll\omega_c$, this expression simplifies to $\Delta_{\rm 1p}/\omega_c \sim \sqrt{\gamma}= g_D \sqrt{1+n}/\omega_c$. Given the analytic eigenvalues and eigenvectors of the truncated Hamiltonian, the Chern number of each band can be analytically evaluated to derive the Ansatz given for the Berry phase in Ref.~\cite{Dag2024}. The gap associated with the exchange of two photons can be approximately derived from the 6-dimensional truncated subspace spanned by photon states $|n\rangle$ and $|n+1\rangle$, $|n+2\rangle$ and by the same electronic states as before~\cite{supp}.

Now we extend the (1-photon exchange) DJC model to a continuum edge theory by imposing an edge at \(y=0\) and replacing \(k_y\!\to\!-i\partial_y\).
For the following formulae, $k_x$ is valley-relative momentum in units of $\omega_c/v_F$, $y$ and $\xi$ are measured in units of $v_F/\omega_c$, and $E$ is measured in units of $\omega_c$ from the DJC model resonance point, $\omega_c/2$.
The resulting generalized eigenvalue problem~\cite{supp}, leads to evanescent solutions
\(\psi(x,y)=e^{ik_xx+\zeta y}\phi\), where $\Re\zeta<0$ for modes localized at the left edge and $\xi^{-1}=|\Re\zeta|$ defines the
inverse
localization length ($\Re\zeta>0$ for the right edge). Note that bulk states have $\Re\zeta=0$. Defining $\tilde{g}=g_D/\omega_c$, the solutions, to leading order in $\tilde{g}$, are
\begin{equation}
\begin{aligned}\label{eq:continuum}
k_{x,0}&=-\frac{\tilde{g}}{2},\ \ E_\pm(k_x)=\pm \frac{\tilde{g}}{1+\tilde{g}^2} \bigl(k_x-k_{x,0}\bigr),\\
{\xi^{-1}(k_x)}&={\tilde g/2-[\tilde g^2/(1+\tilde g^2)](k_x-k_{x,0}).}
\end{aligned}
\end{equation}
where $k_{x,0}$ is the momentum of the crossing point between left and right edge modes, and $E_{\pm}(k_x)$ are the energies of the left and right edge modes, with edge velocity magnitude \(v_{\mathrm{edge}} = v_F\tilde{g}/(1+\tilde{g}^{2})\). All quantities depend on a single energy scale, the light-matter interaction strength \(g_D\). At the crossing, $\xi^{-1}=\tilde{g}/2$, and at weak-coupling this ties the bulk gap to the edge localization \(\Delta_{\rm 1p}\propto\xi^{-1}\).
At the crossing between $n=0$ and $n=1$, the
edge mode has equal photon populations \(p_0=p_1=1/2\), in agreement with the edge mode profile in Fig.~\ref{fig:fig1}(e). See SM for the leading order corrections to the photon populations.

\begin{figure}[!tbp]
\centering
\includegraphics[width=1.00\columnwidth]{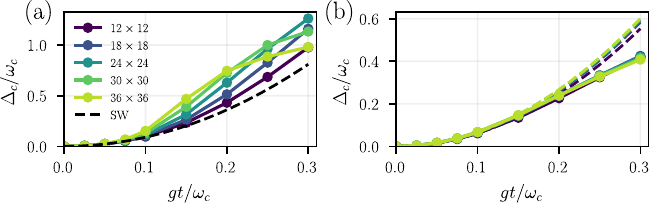}
\caption{\textbf{Finite-density signatures of the cavity-induced gaps.}
Charge gaps $\Delta_c/\omega_c$ versus $gt/\omega_c$, calculated using DMRG on periodic honeycomb lattices of the indicated sizes.
(a) Half filling, $N=L^2/2$, with $\omega_c/t=0.1$ and photon cutoff $n_{\rm ph}=2$.
(b) Near the one-photon resonance, $N=L^2/2+4$, with $\omega_c/t=4\sin(\pi/L)$ and $n_{\rm ph}=1$.
Both panels show a clear opening of charge gaps with increasing light--matter coupling.
Colors identify system sizes in both panels; dashed curves show Schrieffer--Wolff predictions.
Calculation and validation details are given in the SM.}
\label{fig:fig2}
\end{figure}
Fig.~\ref{fig:fig4} benchmarks Eqs.~\eqref{eq:continuum} against the results obtained from the
$T$-matrix method: (a) shows the crossing momentum \(k_{x,0}\), while
(b) compares the inverse
localization length \(\xi\) and (c) the edge velocity
\(v_{\mathrm{edge}}\). We observe that as the light-matter interaction strength increases, the momentum crossing point between two chiral hybrid edge modes moves further away from $\mathbf{K},\mathbf{K'}$ points, localization length decreases as the gap increases and the edge modes acquire a larger dispersion. The agreement between two methods is quantitative up to the
ultrastrong coupling regime~\cite{KonoUltrastrong, kockum2019ultrastrong}, and the deviations at very large \(g\) arise from lattice
features absent in the continuum theory.  This shows that hybrid edge physics is
governed by the DJC model at order \(\mathcal{O}(g_D)\).

\textit{Finite-density signatures.}~To examine whether the cavity-induced gaps persist at finite electron density, we calculate many-body charge gaps using DMRG on system size up to $36\times 36$,
\begin{equation}
\Delta_c=E(N+1)+E(N-1)-2E(N),
\label{eq:dmrg_gap_common}
\end{equation}
where $E(N)$ is the lowest energy in the particle-number sector considered.
At half filling, we find that the charge gap increases quadratically in weak light-matter coupling following the Schrieffer-Wolff (SW) based perturbation theory and single-electron prediction~\cite{Dag2024, supp}. Deviation from quadratic behavior start becoming apparent for $t g/\omega_c > 0.1$, Fig.~\ref{fig:fig2}(a), where we also observe a gap enhancement with electron number. For $t g/\omega_c > 0.2$ however, the gap does not grow monotonically in system size, pointing to a possible gap closing. Nevertheless, the results exhibit a stable charge gap in thermodynamic limit at finite electronic density for some $t g/\omega_c$, where single-electron modeling is still useful. 

{We next tune the cavity frequency to a one-photon resonance, $\epsilon_-(\mathbf k_F)+\omega_c=\epsilon_+(\mathbf k_F)$ in Fig.~\ref{fig:fig1}(b,c), at the chosen electron filling, where the hybrid bright edge modes reside in the single-electron picture. Fig.~\ref{fig:fig2}(b) shows that turning on the cavity coupling opens a robust finite charge gap at this electron filling. 
This many-body gap grows quadratically with $g$ at weak coupling, which can also be understood within SW transformation (dashed), and scales in $g$ less than quadratic as $gt/\omega_c > 0.2$. This charge gap is expected, exactly because of strong light-matter hybridization of different photon copies at a photon exchange gap. Therefore, we conclude that the single-electron picture of edge modes discussed above can persist in finite electronic densities. See SM~\cite{supp} for the details.}

\textit{Conclusions and Outlook}.~We showed that coupling graphene on a ribbon to a chiral cavity gives a semi-metal a tunable chiral edge channel. A change in mirror spacing or alignment, or the circular polarization of the cavity field, can control the light–matter interaction scale \(g_D\), and therefore the edge mode velocity and localization length. The effects of cavity dissipation on the chiral hybrid edge modes is yet to be explored. Crucially, we found many-body charge gaps with DMRG at the gap locations in hybrid band structures, connecting single- and many-electron pictures.

We remark that the Peierls substitution expanded up to the first order in photon operators is controlled, as we focus on weak light-matter interactions. Remaining in this perturbative regime and showing how the hybrid edge modes residing within the photon exchange gaps are localized at the valleys, together with the matching results we observe from the gauge-invariant continuum edge theory assures us about the physical relevance of our numerical results obtained from the tight-binding model.

Chiral vacuum field mediates long-range electron–electron interactions~\cite{PhysRevB.99.235156,Dag2024}, which might, in the long-wavelength limit, renormalize the edge to a cavity-tunable chiral Luttinger liquid~\cite{PhysRevLett.64.2206,jia2022tuning,PhysRevResearch.3.L032013,giamarchi2003quantum}. If this conjecture is true, the plasmon velocity resulting from the Luttinger liquid theory~\cite{giamarchi2003quantum} can be \textit{in situ} tunable with the cavity geometry and light-matter coupling through the edge velocity. Hence this setup might offer a platform to explore strongly correlated 1D physics and quantum transport.

\textit{Acknowledgments}
We thank Martin Claassen, Titus Neupert, Taylor L. Hughes, Julian May-Mann and Takashi Oka for the stimulating discussions.
O.K.D., V.R. and C.B.D acknowledge support from the NSF through a grant for ITAMP at Harvard University. V.R. acknowledges the financial support provided by the Villanova University Summer Grant program during the summer of 2026.

\bibliographystyle{apsrev4-1}
\bibliography{Bibliography}

\clearpage
\appendix
\begin{widetext}
\setcounter{equation}{0}
\setcounter{figure}{0}
\setcounter{page}{1}
\makeatletter
\renewcommand{\theequation}{S\arabic{equation}}
\renewcommand{\thefigure}{S\arabic{figure}}
\renewcommand{\theHfigure}{supp.\arabic{figure}}
\renewcommand{\thetable}{S\arabic{table}}

\section{Supplementary Material}
\subsection{I. Derivation of graphene in a cavity}
\subsubsection{Bulk Hamiltonian}

Here, we present an alternative, and perhaps more widely-used method in the literature for the derivation of the graphene tight-binding Hamiltonian subject to enhanced vacuum fluctuations. Namely, we perform Peierls substitution, which recovers the minimal-coupling Hamiltonian derived in the Ref.~\cite{Dag2024}, when it is expanded to the first order only. The Peierls substitution is justified by the long-wavelength cavity limit: the cavity wavelength is many orders of magnitude larger than the graphene lattice spacing, so the vector potential is uniform on the scale of each nearest-neighbor bond. The subsequent expansion is a small-vacuum-field expansion of this Peierls phase. Let us sketch this derivation here. Graphene has a triangular Bravais lattice with 2-site basis. Hence its Hamiltonian in real space is
\begin{eqnarray}
H(\bm{r}) = t \sum_{\bm{r},n} \left(c^{\dagger}_{\bm{r}} c_{\bm{r+\delta_n}} + \text{h.c.}\right)
\end{eqnarray}
where tunneling occurs between the two sublatices, A and B. This translates to momentum space as
\begin{eqnarray}
H(\bm{k}) = t \sum_{\bm{k},n} \left( e^{i\bm{\delta}_n\cdot \bf{k}} c_{A,\bf{k}}^{\dagger} c_{B,\bf{k}} + \text{h.c.} \right).
\end{eqnarray}
The convention for the distances between the nearest neighbor sites is chosen to be
\begin{align}
\label{nnDistances}
  \boldsymbol{\delta}_1 = \frac{\alpha}{2}\left(\sqrt{3}, 1\right), \,\, \boldsymbol{\delta}_2 = \frac{\alpha}{2}\left(-\sqrt{3}, 1\right), \,\,   \boldsymbol{\delta}_3 =\alpha \left(0,-1\right),
\end{align}
where $\alpha$ is the lattice spacing. We perform a Peierls substitution in the real-space graphene to couple to chiral cavity field,
\begin{eqnarray}
H(\bm{r}) = t\sum_{\bm{r},n} e^{-i g (\hat a\mathbf{e} +\hat a^{\dagger}\mathbf{e}^*)\cdot \bm{\delta}_n}c^{\dagger}_{\bm{r}} c_{\bm{r+\delta_n}}  + \text{h.c.} + \omega\left(\hat a^{\dagger}\hat a+\frac{1}{2}\right),
\end{eqnarray}
where $\mathbf{e} = (1,i)/\sqrt{2}$ is the circular polarization vector of the cavity field and $g$ is the light-matter coupling strength. Then in momentum space, we have
\begin{eqnarray}
H(\bm{k}) =t \sum_{\bm{k},n} e^{-\textrm{i}g (\hat a \mathbf{e} +\hat a^{\dagger} \mathbf{e}^*)\cdot \bm{\delta}_n} e^{i\bm{\delta}_n \cdot \bf{k}} c_{A,\bf{k}}^{\dagger} c_{B,\bf{k}}+ \text{h.c.} + \omega\left(\hat a^{\dagger}\hat a+\frac{1}{2}\right).
\end{eqnarray}
We expand the phase factor due to cavity field up to the first order,
\begin{eqnarray}
H(\bm k) = t\sum_{\bm{k},n} \left( e^{i\bm{\delta}_n\cdot \bf{k}} c_{A,\bf{k}}^{\dagger} c_{B,\bf{k}} - \textrm{i}g (\hat a \mathbf{e} +\hat a^{\dagger}\mathbf{e}^*) \cdot \bm{\delta}_n e^{i\bm{\delta}_n \cdot \bf{k}}  c_{A,\bf{k}}^{\dagger} c_{B,\bf{k}}  + \text{h.c.}\right) + \omega\left(\hat a^{\dagger}\hat a+\frac{1}{2}\right)+\mathcal{O}(g^2)
\end{eqnarray}
The first term is the bare graphene Hamiltonian. The second term includes one-photon virtual processes. Let us now write the matter-photon coupling Hamiltonian in the spinor basis of the sublattices by making a gauge transformation in the graphene Hamiltonian,
\begin{eqnarray}
  H^{\text{pm}}_{\bf{k}} &=& \sum_{\bm{k},n}  \left( -\textrm{i}tg (\hat a\mathbf{e} +\hat a^{\dagger}\mathbf{e}^*) \cdot \bm{\delta}_n e^{i\bm{\delta}_n\cdot \bf{k}}  c_{A,\bf{k}}^{\dagger} c_{B,\bf{k}}  + \text{h.c.}\right) +\mathcal{O}(g^2), \\
  &=& \sum_{\bm{k}} \left( -\textrm{i}tg  e^{i\bm{\delta}_3 \cdot \bf{k} }(\hat a\mathbf{e} +\hat a^{\dagger}\mathbf{e}^*)\cdot \left[ \bm{\delta}_1 e^{i \bm{a}_1\cdot \bf{k}}+\bm{\delta}_2 e^{i \bm{a_2}\cdot \bf{k}}+\bm{\delta}_3 \right] c_{A,\bf{k}}^{\dagger} c_{B,\bf{k}}  + \text{h.c.}\right) +\mathcal{O}(g^2),
\end{eqnarray}
where $\bm{a}_1$ and $\bm{a}_2$ are the primitive vectors of graphene lattice, $\bm{a}_1=\bm{\delta}_1-\bm{\delta_3}$ and $\bm{a}_2=\bm{\delta}_2-\bm{\delta_3}$. Therefore,
\begin{eqnarray}
  H^{\text{pm}}_{\bf{k}} &=& \sum_{\bm{k},n} \bigg( \frac{-\textrm{i}tg}{\sqrt{2}} e^{i\bm{\delta}_3 \cdot \bf{k} } \left[ (\hat a+ \hat a^{\dagger})\delta_{n,x} + i(\hat a-\hat a^{\dagger})\delta_{n,y}\right] \left[\cos(\bm{a}_n \cdot \mathbf{k})+ i \sin(\bm{a}_n \cdot \bf{k})\right] c_{A,\bf{k}}^{\dagger} c_{B,\bf{k}} + \text{h.c.} \bigg) +\mathcal{O}(g^2),\notag
\end{eqnarray}
In the following, the overall phase factor $e^{i\bm{\delta}_3 \cdot \bf{k} }$ is gauged away as $c_{B,\bf{k}} \rightarrow e^{i\bm{\delta}_3 \cdot \bf{k} } c_{B,\bf{k}}$. In this gauge, we use $\bm a_3=0$ and the sublattice spinor $\Psi_{\bm k}=(c_{A,\bm k},c_{B,\bm k})^{\mathsf T}$. The Hermitian photon coordinate on bond $n$ is $(\hat a+\hat a^\dagger)\delta_{n,x}+i(\hat a-\hat a^\dagger)\delta_{n,y}$, consistently with $\mathbf e=(1,i)/\sqrt2$. Converting the hopping and its Hermitian conjugate to Pauli matrices removes the explicit factor $-i$, giving
\begin{eqnarray}
 H^{\text{pm}}_{\bf{k}} &=&
 \frac{tg}{\sqrt{2}}\sum_{\bm{k},n} \left[\hat a (\delta_{n,x}+i\delta_{n,y})+\hat a^{\dagger}(\delta_{n,x}-i\delta_{n,y})\right] \notag \\
 && \times\Psi_{\bm k}^{\dagger}\left[\cos(\bm a_n\cdot\bm k)\sigma_y+\sin(\bm a_n\cdot\bm k)\sigma_x\right]\Psi_{\bm k}+\mathcal O(g^2),
\label{eq:one-polarization-fourier}
\end{eqnarray}
which is also what has been obtained with minimal substitution in Ref.~\cite{Dag2024}.

\subsubsection{Graphene Hamiltonian on zig-zag strip}
The Hamiltonian for graphene on a zigzag strip follows as \begin{eqnarray}
 H_{\rm 0} = t\sum_{k_x}\sum_{j=1}^{N_y/2} \biggl[\left(c_{2j,k_x}^{\dagger}c_{2j+1,k_x} + \rm h.c.\right)
+\biggl\{c_{2j,k_x}^{\dagger}c_{2j-1,k_x} \left(e^{i k_x \alpha \sqrt{3}/2}+
e^{-i k_x \alpha \sqrt{3} /2}\right) + \rm h.c.\biggr\}\biggr],\label{eq:circ_real2}
\end{eqnarray}
 where $N_y$ is the number of atomic rows and $j=1,\ldots,N_y/2$ labels zigzag unit cells. The operators $c_{j,k_x}$ annihilate fermions at position $j$ along the strip with momentum $k_x$ in the orthogonal direction and obey $\lbrace c^{\dagger}_{j,k_x},c_{s,k_x'} \rbrace = \delta_{k_xk_x'}\delta_{js}$. In the following we simplify the notation for the momentum with $k_x \equiv k$. When the cavity field is coupled via the Peierls substitution,
\begin{eqnarray}
\mathcal{H} &=& t\sum_{j,k} \biggl[\left(e^{-\textrm{i}g (\hat a\mathbf{e}+\hat a^{\dagger}\mathbf{e}^*)\cdot \mathbf{\delta}_3} c_{2j,k}^{\dagger}c_{2j+1,k} + \rm h.c.\right) \notag \\
&+&\biggl\{c_{2j,k}^{\dagger}c_{2j-1,k} \left(e^{-\textrm{i}g (\hat a\mathbf{e}+\hat a^{\dagger}\mathbf{e}^*)\cdot \mathbf{\delta}_1} e^{i k \alpha \sqrt{3}/2}+
e^{-\textrm{i}g (\hat a\mathbf{e}+\hat a^{\dagger}\mathbf{e}^*)\cdot \mathbf{\delta}_2}e^{-i k \alpha \sqrt{3} /2}\right) + \textrm{h.c.}\biggr\}\biggr] +\omega\left(\hat a^{\dagger}\hat a+\frac{1}{2}\right).
\end{eqnarray}
To make this expression linear in vector potential, let us expand the Peierls phase,
\begin{eqnarray}
 \mathcal{H} &=& H_{\rm 0} -\textrm{i}tg (\hat a\mathbf{e}+\hat a^{\dagger}\mathbf{e}^*) \cdot  \sum_{j,k} \biggl[ \bm \delta_3 c_{2j,k}^{\dagger}c_{2j+1,k} + c_{2j,k}^{\dagger}c_{2j-1,k}  \left(\bm\delta_1 e^{i k \alpha \sqrt{3}/2}+
\bm \delta_2 e^{-i k \alpha \sqrt{3} /2}\right)\biggr]+ \textrm{h.c.} +\omega\left(\hat a^{\dagger}\hat a+\frac{1}{2}\right) \notag
\end{eqnarray}
With right handed circular polarization, this expression reduces to
\begin{eqnarray}
\mathcal{H} &=& H_{\rm 0} -\textrm{i} \frac{t g}{\sqrt{2}} \sum_{j,k} \biggl[ \left[ (a+a^{\dagger})\delta_{3,x} + i(a-a^{\dagger})\delta_{3,y}\right] c_{2j,k}^{\dagger}c_{2j+1,k}   \\
&+& c_{2j,k}^{\dagger}c_{2j-1,k} \left(\left[ (a+a^{\dagger})\delta_{1,x} + i(a-a^{\dagger})\delta_{1,y}\right] e^{i k \alpha \sqrt{3}/2}+
\left[ (a+a^{\dagger})\delta_{2,x} + i(a-a^{\dagger})\delta_{2,y}\right]e^{-i k \alpha \sqrt{3} /2}\right)\biggr] + \rm h.c.\notag \\
&+& \omega\left(\hat a^{\dagger}\hat a+\frac{1}{2}\right).\notag
\end{eqnarray}
This completes the derivation of Eq.~\eqref{eq:circ_real} in the main text. Note that the main text normalizes momentum $k$ with lattice constant $\alpha$, or in other words $\alpha=1$ is set.

\subsection{II. Semi-infinite methods and T-matrix}
Using Eq.~\eqref{eq:circ_real2}, we compute the spectrum with open boundaries by exact diagonalization and compare it to the infinite system with periodic boundary conditions. Finite ribbon width limits the resolution of transverse bulk modes, with a momentum scale of order the inverse width. It does not limit the sampling of the conserved edge momentum $k_x$, nor does it by itself exclude overlap with bulk states at the same energy and $k_x$. Resolving weakly localized boundary states can nevertheless require a very wide ribbon. A complementary route that avoids large real-space calculations is the T-matrix approach~\cite{PhysRevB.48.11851,PhysRevB.93.134304}. To this end, we work in the single-electron basis
\begin{equation}
|\psi(k_x)\rangle
= \sum_{n=0}^{N_\mathrm{ph}}\sum_{j=1}^{N_y}\psi_{n j}(k_x)\,
\hat{c}^\dagger_{j,k_x}\,\frac{(\hat a^{\dagger})^n}{\sqrt{n!}}\,|\Omega\rangle,
\end{equation}
where $j$ labels the $j$-th lattice site along $y$, $k_x$ is the plane-wave momentum along $x$, $n$ is the photon number, and $|\Omega\rangle$ is the joint vacuum of electrons and photons. Consider the Schr\"odinger equation
$\langle n,j,k_x|\hat{H}|\psi(k_x)\rangle = E\,\psi_{n,j}(k_x)$. In the following, we treat $k_x$ as fixed and suppress it in the notation. In this basis the Hamiltonian is periodic under $j\to j+2$, and the unit cell has dimension
$d=2(N_\mathrm{ph}+1)\equiv 2r$.
We therefore relabel the wave function using $l=\lfloor j/2\rfloor$ and (pseudo)spinor indices
$\sigma = 2n + (j \bmod 2)$.
The Schr\"odinger equation then takes the form~\cite{PhysRevB.48.11851}
\begin{equation}
    \sum_{\sigma'} J^\dagger_{\sigma\sigma'}\, \psi_{l-1,\sigma'} +
   \bigl(M_{\sigma\sigma'}-\delta_{\sigma\sigma'}E\bigr)\,\psi_{l,\sigma'}
    +  J_{\sigma\sigma'}\,\psi_{l+1,\sigma'} = 0,
   \label{eq:schrodinger}
\end{equation}
with nilpotent\footnote{We choose a unit cell large enough that $J$ is nilpotent, $J^2=0$. If $d$ is insufficient, we can, for example, take $d\to 2d$.} hopping matrices $J,J^\dagger$ between unit cells ($J$ is the block $H_{l,l+1}$) and a Hermitian matrix $M$ within the unit cell, which defines the (single-electron) Green’s function
\begin{equation}
   G_0 = \bigl[E\,\mathbb{I} - M\bigr]^{-1},
   \label{eq:G0}
\end{equation}
well defined unless $E\,\mathbb{I}-M$ is singular.\footnote{Singularity occurs at band centers; in that case, one eigenvalue of $M$ equals $E$, and we introduce an infinitesimal $i\epsilon$ to render the inverse well defined.} The matrices $J$ and $J^\dagger$ are non-normal and thus lack a common set of orthonormal eigenvectors, but they admit a singular-value decomposition
\begin{equation}
   J = V\,\Xi\,W^\dagger,
\end{equation}
where we keep only the $r$ nonzero singular values in $\Xi$. Thus $V,W$ are $d\times r$ matrices with $V^\dagger V=W^\dagger W=\mathbb I_r$ Nilpotency implies
\begin{equation}
   V^\dagger W = W^\dagger V = 0,
   \label{eq:nullpotent}
\end{equation}
so the selected column spaces of $V$ and $W$ are mutually orthogonal.
Any state in the $d$-dimensional spinor space decomposes as
\begin{equation}
    \psi_\sigma = \sum_{\sigma'=1}^r \bigl(V_{\sigma\sigma'}\alpha_{\sigma'} + W_{\sigma\sigma'}\beta_{\sigma'}\bigr)
    \quad\Leftrightarrow\quad
    \bar{\psi} = V\,\bar{\alpha} + W\,\bar{\beta},
    \label{eq:decomp}
\end{equation}
with coefficients $\bar{\alpha},\bar{\beta}\in\mathbb{C}^r$. Using Eq.~\eqref{eq:nullpotent},
\begin{align}
    J\,\bar{\psi} = V\,\Xi\,\bar{\beta}, \qquad
    J^\dagger\,\bar{\psi} = W\,\Xi\,\bar{\alpha}.
    \label{eq:Jaction}
\end{align}
Combining Eqs.~\eqref{eq:G0}, \eqref{eq:decomp}, and \eqref{eq:Jaction}, Eq.~\eqref{eq:schrodinger} becomes
\begin{equation}
   V\,\bar{\alpha}_l + W\,\bar{\beta}_l
   = G_0\,W\,\Xi\,\bar{\alpha}_{l-1} + G_0\,V\,\Xi\,\bar{\beta}_{l+1}.
\end{equation}
Since $V$ and $W$ are orthogonal, multiplying by $V^\dagger$ and $W^\dagger$ yields
\begin{equation}
\begin{aligned}
\bar{\alpha}_l &= V^\dagger G_0 W\,\Xi\,\bar{\alpha}_{l-1} + V^\dagger G_0 V\,\Xi\,\bar{\beta}_{l+1},\\
\bar{\beta}_l  &= W^\dagger G_0 W\,\Xi\,\bar{\alpha}_{l-1} + W^\dagger G_0 V\,\Xi\,\bar{\beta}_{l+1},
\end{aligned}
    \label{eq:schrod2}
\end{equation}
which motivates the definition of in/out-channel Green’s functions
$\mathcal{G}_{ab} \equiv A^\dagger G_0 B\,\Xi$ with $a,b\in\{v,w\}$ and the corresponding matrices $A,B\in\{V,W\}$.
Equation~\eqref{eq:schrod2} can then be written as\,(this form differs from Ref.~\cite{PhysRevB.48.11851}, which contains a typographical error in the analogous equation)
\begin{equation}
\begin{aligned}
   \underbrace{\begin{pmatrix}
      \mathcal{G}^{-1}_{wv} & -\mathcal{G}^{-1}_{wv}\mathcal{G}_{ww} \\
      \mathcal{G}_{vv}\mathcal{G}^{-1}_{wv} & \mathcal{G}_{vw} - \mathcal{G}_{vv}\mathcal{G}^{-1}_{wv}\mathcal{G}_{ww}
   \end{pmatrix} }_{\equiv\, T(E)}
   \begin{pmatrix} \bar{\beta}_l \\ \bar{\alpha}_{l-1}\end{pmatrix}
   &= {\begin{pmatrix} \bar{\beta}_{l+1}\\ \bar{\alpha}_l\end{pmatrix}}
   \\
   \Leftrightarrow\quad T(E)\,\bar{\phi}_l &= \bar{\phi}_{l+1}.
\end{aligned}
\end{equation}
The (generally non-normal) transfer matrix \(T(E)\) acts on the channel vector \(\bar\phi_l=(\bar\beta_l,\bar\alpha_{l-1})^{\mathsf T}\), not directly on the physical cell spinor \(\bar\psi_l\).  For an infinite system, Bloch's theorem gives \(\bar\phi_{l+1}=\lambda\bar\phi_l\) with \(\lambda=e^{ik_y}\).  Right and left eigenvector matrices \(\mathcal R\) and \(\mathcal L\) are defined by
{
\begin{equation}
\begin{aligned}
    T\mathcal R&=\mathcal R\Lambda,\\
    \mathcal L^\dagger T&=\Lambda\mathcal L^\dagger,\\
    \mathcal L^\dagger\mathcal R&=\mathbb I,
\end{aligned}
\label{eq:tm_biorthogonal}
\end{equation}
}
where \(\Lambda=\mathrm{diag}(\lambda_s)\).  Boundary-compatible finite or semi-infinite solutions are superpositions of these transfer eigenvectors.
The magnitude of $\lambda$ determines the character of the state: $|\lambda_s|<1$ gives a left-edge state, $|\lambda_s|>1$ a right-edge state, and $|\lambda_s|=1$ a standing wave. We partition the eigenvectors accordingly: growing $(|\lambda|>1)$, decaying $(|\lambda|<1)$, and constant magnitude $(|\lambda|=1)$, denoted $\mathcal{R}_>$, $\mathcal{R}_<$, and $\mathcal{R}_0$, respectively. For the nonzero eigenvalues, the associated (generally oblique) projectors are
\begin{equation}
\begin{aligned}
   \mathcal{P}_> &= \mathcal{R}_>\,(\mathcal{L}_>^\dagger \mathcal{R}_>)^{-1}\,\mathcal{L}^\dagger_>,\\
   \mathcal{P}_< &= \mathcal{R}_<\,(\mathcal{L}_<^\dagger \mathcal{R}_<)^{-1}\,\mathcal{L}^\dagger_<.
\end{aligned}
\label{eq:tm_oblique_projectors}
\end{equation}

For a transfer eigenvector \(u_s=(\beta_s,\alpha_s)^{\mathsf T}\) defined at a reference cell, the physical spinor in cell \(l\) must be reconstructed after one transfer step.  At that reference cell, \(\alpha_l=\lambda_s\alpha_s\) and \(\beta_l=\beta_s\), so
{
\begin{equation}
\psi_l^{(s)}=\lambda_s^l\chi_s,
\qquad
\chi_s=W\beta_s+\lambda_sV\alpha_s.
\label{eq:tm_physical_spinor}
\end{equation}
}
Thus an edge state is \(\psi_l=\sum_s c_s\lambda_s^l\chi_s\), with the coefficients \(c_s\) fixed by the boundary condition.  A local physical-cell observable is evaluated directly from the reconstructed spinor,
{
\begin{equation}
\langle O\rangle_l=
\frac{\psi_l^\dagger O\psi_l}{\psi_l^\dagger\psi_l}.
\label{eq:tm_local_observable}
\end{equation}
}
For the normalized edge state, the corresponding integrated observable is
{
\begin{equation}
\langle O\rangle_{\rm edge}=
\frac{\displaystyle\sum_l\psi_l^\dagger O\psi_l}
{\displaystyle\sum_l\psi_l^\dagger\psi_l}.
\label{eq:tm_edge_observable}
\end{equation}
}
In particular, photon occupations are defined by the physical-cell photon-number operator in these expressions.
We therefore define the positive inverse amplitude localization length as
\begin{equation}
{\xi^{-1}=\min_{s\in\mathrm{edge}}\left|\log\!\bigl(|\lambda_s|\bigr)\right|,}
\label{eq:tm_xi}
\end{equation}
where the minimum is taken over the transfer-matrix components participating in the chosen edge state; for a single dominant component this reduces to $\xi=1/\left|\log|\lambda_\mathrm{dom}|\right|$. This length is measured in transfer cells. One step advances two atomic rows, whose alternating separations are $\alpha/2$ and $\alpha$; hence the physical cell size is $3\alpha/2$, and the physical inverse length is $\xi^{-1}/(3\alpha/2)$. The profile in Fig.~\ref{fig:fig4}(d) instead uses the atomic-row index $y/d_y$, which should not be identified with the transfer-cell coordinate. For fully gapped edge modes localized at either boundary, we impose
\begin{equation}
   \mathcal{P}_R =
   \begin{pmatrix}
     0_r & 0_r\\
     0_r & \mathbb{I}_{r}
   \end{pmatrix},
   \qquad
   \mathcal{P}_L =
   \begin{pmatrix}
     \mathbb{I}_{r} & 0_r\\
     0_r & 0_r
   \end{pmatrix}.
\label{eq:tm_boundary_projectors}
\end{equation}
A transfer vector $\bar{\phi}$ that solves the Schr\"odinger equation and satisfies the boundary conditions thus obeys, for the left edge,
\begin{equation}
    {\mathcal{P}_<\,\bar{\phi}_l = \bar{\phi}_l \quad\text{and}\quad \mathcal{P}_L\,\bar{\phi}_l = \bar{\phi}_l},
   \label{eq:cond1}
\end{equation}
and, for the right edge,
\begin{equation}
    {\mathcal{P}_>\,\bar{\phi}_l = \bar{\phi}_l \quad\text{and}\quad \mathcal{P}_R\,\bar{\phi}_l = \bar{\phi}_l}.
   \label{eq:cond2}
\end{equation}
Note that $\mathcal{P}_<$ and $\mathcal{P}_L$ (and, analogously, $\mathcal{P}_>$ and $\mathcal{P}_R$) generally do not commute. Moreover, eigenspace projectors for non-normal matrices are themselves typically non-normal and therefore need not define orthogonal subspaces. Nevertheless, as projectors ($\mathcal{P}^2=\mathcal{P}$) they are diagonalizable with eigenvalues $0$ and $1$. Common eigenvectors of the relevant projectors furnish the desired edge-state solutions at energy $E$.
To isolate candidate branches associated with lower photon-exchange gaps~\cite{wielian2025transfer}, we also use a channel selector,
\[
\mathcal{B}=(1,\ldots,1,\underbrace{0}_{i_1},1,\ldots,1,\underbrace{0}_{i_2,\dots},\ldots,1,1),
\]
where the indices $i_1,\ldots,i_n$ identify entries of the transfer-channel vector $\bar\phi$, not directly physical photon populations. Their physical content follows from the reconstruction in Eq.~\eqref{eq:tm_physical_spinor}. The corresponding projector is $\mathcal P_\mathcal B=\mathrm{diag}[\mathcal B]$. Such a selector filters candidate branches; it does not replace the physical open-boundary condition. In the convention of Eq.~\eqref{eq:schrodinger}, an admissible left-edge state must satisfy
{
\[
\begin{aligned}
(M-E\mathbb I)\psi_0+J\psi_1&=0,\\
J^\dagger\psi_{l-1}+(M-E\mathbb I)\psi_l+J\psi_{l+1}&=0\qquad(l\geq1).
\end{aligned}
\]
}
The decaying-subspace condition and the selected channel indices alone do not establish these physical equations.

\subsection{III. Derivation of the Dirac-Jaynes-Cummings model}

We derive the Dirac-Jaynes-Cummings model from the tight-binding Hamiltonian.
With \(\Psi_{\mathbf k}=(c_{A,\mathbf k},c_{B,\mathbf k})^{T}\) in the sublattice basis, the Fourier transform gives
\begin{equation}
H_0(\mathbf k)=
t\,
\Psi_{\mathbf k}^{\dagger}
\begin{pmatrix}
0 & f(\mathbf k) \\ f^{*}(\mathbf k) & 0
\end{pmatrix}
\Psi_{\mathbf k},
\qquad
f(\mathbf k)=\sum_{n=1}^{3}e^{i\mathbf k\!\cdot\!\boldsymbol\delta_n},
\label{eq:fourier_tb}
\end{equation}
where the bond vectors are
\(\boldsymbol\delta_{1,2}=\tfrac{\alpha}{2}(\pm\sqrt3,1)\) and
\(\boldsymbol\delta_{3}=\alpha(0,-1)\), as before. Near the valley point \(\mathbf K=(4\pi/(3\sqrt3\alpha),0)\) we set \(\mathbf k=\mathbf K+\mathbf q\) with \(|\mathbf q|\alpha\ll1\). Expanding \(f(\mathbf k)\) to linear order and measuring momenta in units of $\alpha^{-1}$ yields
\begin{equation}
{f(\mathbf K+\mathbf q)\simeq
-\frac{3}{2}\bigl(q_x-iq_y\bigr)},
\label{eq:linear_f}
\end{equation}
so that the free part assumes the massless-Dirac form
\begin{equation}
H_e=
v_F\,
\Psi_{\mathbf q}^{\dagger}
\bigl(
  q_x\sigma_x + q_y\sigma_y
\bigr)
\Psi_{\mathbf q},
\label{eq:dirac_free}
\end{equation}
with Pauli matrices \(\sigma_{x,y}\) acting in the sublattice space and the Fermi velocity \(v_F=\tfrac{3}{2}\,t\). Next we linearize the Peierls interaction. Substituting the vector potential into the second term of Eq.~\eqref{eq:circ_real}, performing the Bloch transform, and again expanding to lowest order in \(\mathbf q\) (while keeping \(\mathbf A\) to zeroth order) gives
\begin{equation}
H_{\text{int}}=
{\frac{g_D}{2}}\,
\Psi_{\mathbf q}^{\dagger}
\Bigl[ \bigl(\sigma_x+i\sigma_y\bigr) a +
  \bigl(\sigma_x-i\sigma_y\bigr) a^{\dagger}
\Bigr]
\Psi_{\mathbf q}\,,
\qquad
{g_D=\sqrt2\,v_F g= \frac{3}{\sqrt{2}} t g},
\label{eq:interaction_k}
\end{equation}
where \(\sigma_x\!\pm\!i\sigma_y=2\sigma_{\pm}\) are the sublattice raising and lowering operators. Importantly, at this leading order \(g_D\) is momentum independent. Collecting Eqs.~\eqref{eq:dirac_free}, \eqref{eq:interaction_k}, and the cavity energy \(\omega_c a^{\dagger}a\), the low-energy Hamiltonian becomes
\begin{equation}
H_{\text{DJC}}
=
v_F\,\mathbf{q}\!\cdot\!\boldsymbol{\sigma}
\;+\;
\omega_c\,a^{\dagger}a
\;+\;
g_D
\bigl(
  \sigma^{+}a + \sigma^{-}a^{\dagger}
\bigr),
\label{eq:dirac_jc_final}
\end{equation}
which is a Dirac-Jaynes-Cummings model whose vacuum Rabi frequency is the momentum-independent constant \(g_D\) inherited directly from the lattice Peierls coupling \(g\). Equation~\eqref{eq:dirac_jc_final} provides the analytical backbone for the edge-state wave functions, localization lengths, and time-evolution protocols discussed in the main text.

\subsection{IV. Bulk Dirac-Jaynes-Cummings model}

We first derive the locations of the photon-exchange gap openings, where photons are exchanged between the bands. These occur away from the $K$ and $K'$ points. For simplicity, we relabel the momentum $\mathbf q$ used in the previous section as $\mathbf k$. Starting from Eq.~\eqref{eq:dirac_jc_final}, we take a wave-function ansatz that includes single-photon transitions between $|n\rangle$ and $|n{+}1\rangle$,
\begin{align}
  |\psi_{\mathbf k}\rangle
  = a_{\mathbf k}\,|\downarrow\rangle_{\mathbf k}\,|n\rangle
  + b_{\mathbf k}\,|\uparrow\rangle_{\mathbf k}\,|n\rangle
  + c_{\mathbf k}\,|\downarrow\rangle_{\mathbf k}\,|n{+}1\rangle
  + d_{\mathbf k}\,|\uparrow\rangle_{\mathbf k}\,|n{+}1\rangle .
\end{align}
Acting with the Hamiltonian on this ansatz yields the truncated Hamiltonian
\begin{align}
  H_\mathrm{tr}=
  \begin{pmatrix}
\omega_c \, n & v_F(k_x+i k_y) & 0 & g_D\sqrt{n+1}\\
v_F(k_x-i k_y) & \omega_c \, n  & 0&0\\
0 &0 &\omega_c (n+1) & v_F(k_x+i k_y)\\
g_D\sqrt{n+1} & 0& v_F(k_x-i k_y)&\omega_c (n+1)
\end{pmatrix}.
\label{Eq:EVequation}
\end{align}
The bare resonance momentum is obtained from the uncoupled crossing condition
\begin{align}
    v_F k + n \omega_c &= -\,v_F k + (n+1)\omega_c \\
    \Leftrightarrow\quad
    2v_F k &= \omega_c \\
    \Leftrightarrow\quad
    k &= \frac{\omega_c}{2 v_F}.
\end{align}
At this bare resonance momentum, the splitting of the truncated matrix is
\begin{align}
    \frac{\Delta}{\omega_c}(\chi)=\sqrt{2}\,\sqrt{1+\chi-\sqrt{\,1+\chi+\chi^2\,}},
\end{align}
with $\chi=(g_D/\omega_c)^2(1+n)$. For $\chi\ll1$, $\Delta/\omega_c(\chi)\sim \sqrt{\chi}=g_D\sqrt{1+n}/\omega_c$. The eigenvectors also determine the Berry curvature. For a lattice band, the integer Chern number requires the full Brillouin-zone integral~\cite{bernevig2013topological},
\begin{align}
    C_n=\frac{1}{2\pi i}\int_{\mathcal{BZ}} {d^2\mathbf{k}}\, F_n(\mathbf{k}),
    \label{eq:Cn}
\end{align}
where the Berry curvature of the $n$th band is
\begin{align}
\label{eq:berry_curvature_definition}
    F_n(\mathbf{k})=
    \Big[\partial_{k_x}
    \big\langle \hat{\mathbf{v}}_n(\mathbf{k})\big|\partial_{k_y}\hat{\mathbf{v}}_n(\mathbf{k})\big\rangle
    -\partial_{k_y}
    \big\langle \hat{\mathbf{v}}_n(\mathbf{k})\big|\partial_{k_x}\hat{\mathbf{v}}_n(\mathbf{k})\big\rangle
    \Big],
\end{align}
and $\hat{\mathbf{v}}_n$ is the normalized eigenvector of the $n$th band. Rewriting Eq.~\eqref{Eq:EVequation} by expressing the complex momentum in polar form, $k_x+i k_y = k\,e^{i\theta}$, and rescaling momenta in units of $\omega_c/v_F$ so that $k\,v_F/\omega_c \to k$ and $\tilde{g}=g_D/\omega_c$, we obtain
\begin{align}
\mathcal{H}_\text{tr}/\omega_c =
       \begin{pmatrix}
n & k\,e^{i\theta} & 0 & \tilde{g}\sqrt{n+1}\\
k\,e^{-i\theta} &  n  & 0&0\\
0 &0 &(n+1) & k\,e^{i\theta}\\
\tilde{g}\sqrt{n+1} & 0& k\,e^{-i\theta}& (n+1)
\end{pmatrix}.
\end{align}
The normalized eigenvectors $\hat{\mathbf v}_n$ are obtained analytically by diagonalizing Eq.~\eqref{Eq:EVequation}. Using the polar-derivative identities
\begin{align}
    \partial_{k_x}&=\cos\theta\,\partial_{k}-\frac{\sin\theta}{k}\,\partial_{\theta},\\
    \partial_{k_y}&=\sin\theta\,\partial_{k}+\frac{\cos\theta}{k}\,\partial_{\theta},
\end{align}
Eq.~\eqref{eq:berry_curvature_definition} simplifies to
\begin{align}
    F_n(\mathbf{k})=\frac{2i}{k}\,\mathrm{Im}\!\left(\big\langle\partial_k \hat{\mathbf{v}}_n \big| \partial_{\theta}\hat{\mathbf{v}}_n\big\rangle\right),
\end{align}
which facilitates the evaluation of the valley contribution to the Berry-curvature integral. In particular, for the second hybrid band the valley calculation gives approximately $3/2$, consistent with the lattice value $C=3$ when both valleys are taken into account~\cite{Dag2024}. The integer band invariant nevertheless remains the full-Brillouin-zone lattice result.

\subsubsection{2-photon exchange gap}
The 2 photon gap opening occurs at $k=\omega_c/v_F$. We now aim to determine its gap size. To this end, we extend our wavefunction Ansatz to also allow for states with photon number $|n+2\rangle$:
\begin{align}
   |\psi_k\rangle =a_k|\downarrow\rangle_k |n\rangle+b_k|\uparrow\rangle_k |n\rangle+c_k|\downarrow\rangle_k |n+1\rangle+d_k|\uparrow\rangle_k |n+1\rangle+e_k|\downarrow\rangle_k|n+2\rangle+f_k|\uparrow\rangle_k|n+2\rangle
\end{align}
We use the rescaled momentum of the preceding subsection and measure energy in units of $\omega_c$. Before expanding the characteristic polynomial, we subtract $(n+1)\omega_c$, the bare two-photon resonance center of $H_{\mathrm{DJC}}$; including the cavity zero-point energy would give the center $(n+3/2)\omega_c$. Hence $E$ below is the energy relative to this center. With $k_y=0$, the shifted eigenvalue equation reads
\begin{align}
\left|     \begin{pmatrix}
 {-1} & k_x & 0 & \tilde{g}\sqrt{n+1}&0&0\\
 k_x & {-1}& 0&0&0&0\\
 0 &0 &{0} &k_x&0&\tilde{g}\sqrt{n+2}\\
  \tilde{g}\sqrt{n+1} & 0& k_x&{0}&0&0\\
  0&0&0&0&{1}&k_x\\
  0&0&\tilde{g}\sqrt{n+2}&0&k_x&{1}
\end{pmatrix}-E \mathbbm 1  \right|=0.
\end{align}
At the bare resonance $k_x=1$, its characteristic polynomial is
{
\[
\begin{aligned}
0={}&E^6-[5+(2n+3)\tilde g^2]E^4-\tilde g^2E^3\\
&+[4+2(2n+3)\tilde g^2+(n+1)(n+2)\tilde g^4]E^2
-(n+1)(n+2)\tilde g^4.
\end{aligned}
\]
}
The near-resonance roots have $E=O(\tilde g^2)$. Keeping the constant term and the coefficient of $E^2$ through order $\tilde g^2$ gives their approximate symmetric splitting. Restoring units, the full avoided-crossing gap in this approximation is

\begin{align}
{\frac{\Delta_{2p}}{\omega_c}\simeq\frac{2\tilde{g}^2\sqrt{(n+1)(n+2)}}{\sqrt{4+2\tilde{g}^2(3+2n)}}.}
\end{align}

\subsection{V. Semi-infinite Dirac-Jaynes-Cummings model}

Here we extend the formalism to the edge. We continue to write the rescaled variables as $k_x$, $E$, and $y$: momentum is valley relative and measured in units of $\omega_c/v_F$, while $y$ and $\xi$ are measured in units of $v_F/\omega_c$. We now subtract the vacuum--one-photon resonance center $\omega_c/2$ of $H_{\mathrm{DJC}}$ (or $\omega_c$ when the cavity zero-point energy is included) and divide energies by $\omega_c$. Starting from Eq.~\eqref{Eq:EVequation} in the ordered basis
\(
(|\downarrow,n\rangle,|\uparrow,n\rangle,|\downarrow,n+1\rangle,|\uparrow,n+1\rangle)
\)
and replacing \(k_y\to -i\partial_y\) with a complex wave number \(\zeta\in\mathbb{C}\) whose real part controls the amplitude localization length \(\xi^{-1}=|\Re\zeta|\), the spinor \(\psi\) at energy \(E\) satisfies
\begin{equation}
{ (M-E\mathbb{I}-\zeta B)\psi=0,}
\label{eq:djc_edge_generalized}
\end{equation}
with matrices
\begin{equation}
{
M=\begin{pmatrix}
-\tfrac12 & k_x & 0 & \tilde{g}\\
k_x & -\tfrac12 & 0 & 0\\
0 & 0 & \tfrac12 & k_x\\
\tilde{g} & 0 & k_x & \tfrac12
\end{pmatrix},
\qquad
B=\begin{pmatrix}
0&1&0&0\\ -1&0&0&0\\ 0&0&0&1\\ 0&0&-1&0
\end{pmatrix}.}
\label{eq:djc_edge_matrices}
\end{equation}
For the explicit edge-state formulas below we take the vacuum-one-photon sector \(n=0\). The corresponding formulas for a general adjacent pair \(n,n+1\) follow by replacing \(\tilde g\) with the effective coupling \(\tilde g\sqrt{n+1}\) and centering energy at $(n+1/2)\omega_c$ for $H_{\mathrm{DJC}}$, or $(n+1)\omega_c$ when the cavity zero-point energy is included.
Since \(B^{2}=-\mathbb{I}_{4}\), the determinant factorizes.  Writing \(C(\zeta)=M-E\mathbb{I}-\zeta B\) in \(2\times2\) blocks,
\begin{equation}
{
C=\begin{pmatrix}A & D\\[4pt] D^{\intercal} & \tilde A\end{pmatrix},
\quad
A=\begin{pmatrix}
-\tfrac12-E & k_x-\zeta\\[2pt]
k_x+\zeta & -\tfrac12-E
\end{pmatrix},
\quad
D=\begin{pmatrix}0&\tilde{g}\\ 0&0\end{pmatrix},
\quad
\tilde A=\begin{pmatrix}
\tfrac12-E & k_x-\zeta\\[2pt]
k_x+\zeta & \tfrac12-E
\end{pmatrix}.}
\label{eq:djc_edge_C}
\end{equation}
Evaluating \(\det C\) gives
\begin{equation}
{
\det C=(\zeta^{2}-T-S)\,(\zeta^{2}-T+S),}
\label{eq:djc_edge_determinant}
\end{equation}
with
\begin{equation}
{
T=k_x^{2}-E^{2}-\frac14,\qquad
S^{2}=(1+\tilde{g}^{2})E^{2}-\frac{\tilde{g}^{2}}{4},
\qquad
\zeta^{2}=T\pm S.}
\label{eq:djc_edge_roots}
\end{equation}
Define \(r_{\pm}=\sqrt{T\pm S}\) with \(\Re r_{\pm}>0\); near the node and for \(\tilde g>0\), we choose \(S=+i\tilde g/2\) at \(E=0\).  The four roots are
\(\zeta_{\pm}^{\sigma}=(-1)^{\sigma}r_{\pm}\) for \(\sigma=1,2\), with
\(\zeta_{\pm}^{1}=-r_{\pm}\) (left-decaying) and \(\zeta_{\pm}^{2}=+r_{\pm}\) (right-decaying).
For each root the null vector of \(C(\zeta)\) is
\begin{equation}
{
\phi_{s}(\zeta_{s})=
\begin{pmatrix}
(E+\tfrac12)\tilde{g}\\[2pt]
(k_x+\zeta_{s})\tilde{g}\\[2pt]
-\,\dfrac{(k_x-\zeta_{s})(E+sS)}{\tfrac12-E}\\[6pt]
E+sS
\end{pmatrix},
\qquad s=\pm,}
\label{eq:djc_edge_spinor}
\end{equation}
up to normalization.  The fourth component is \(D(\zeta_{s})=E+sS\).
A physical left-edge state is a superposition of the two left-decaying modes,
\begin{equation}
{
\Psi_{L}(y)=\mathcal{N}\bigl[e^{\zeta_{+}y}\phi_{+}-e^{\zeta_{-}y}\phi_{-}\bigr],}
\label{eq:djc_edge_left_state}
\end{equation}
where the coefficients are chosen so that the first and third components vanish at the boundary \(y=0\).  Imposing this condition yields the determinantal equation
\begin{equation}
{
F_{L}(E,k_x)=(E+S)r_{+}-(E-S)r_{-}+2Sk_x=0.}
\label{eq:djc_edge_left_boundary}
\end{equation}
For the right edge the analogous condition is
\begin{equation}
{
F_{R}(E,k_x)=(E-S)r_{+}-(E+S)r_{-}-2Sk_x=0,}
\label{eq:djc_edge_right_boundary}
\end{equation}
and the two functions are related by
\begin{equation}
{ F_{R}(E,k_x)=-F_{L}(-E,k_x).}
\label{eq:djc_edge_boundary_symmetry}
\end{equation}
At the node
\begin{equation}
{ E_{0}=0,\quad k_{x0}=-\tilde{g}/2,}
\label{eq:djc_edge_node}
\end{equation}
both \(F_{L,R}\) vanish.  Set \(\delta=k_x+\tilde{g}/2\).  Expanding \(F_{L}=0\) gives
\begin{equation}
{
E_{L}= -\frac{\tilde{g}}{1+\tilde{g}^{2}}\,\delta
       +\frac{\tilde{g}^{2}(2-\tilde{g}^{2})}{(1+\tilde{g}^{2})^{3}}\,\delta^{2}
       +\mathcal{O}(\delta^{3}),}
\label{eq:djc_edge_left_dispersion}
\end{equation}
while the right-edge dispersion follows from the symmetry relation,
\begin{equation}
{
E_{R}= -E_{L}
      =\frac{\tilde{g}}{1+\tilde{g}^{2}}\,\delta
       -\frac{\tilde{g}^{2}(2-\tilde{g}^{2})}{(1+\tilde{g}^{2})^{3}}\,\delta^{2}
       +\mathcal{O}(\delta^{3}).}
\label{eq:djc_edge_right_dispersion}
\end{equation}
The edge velocity at the node is therefore
\begin{equation}
{ v_{\mathrm{edge}}=\tilde{g}/(1+\tilde{g}^{2}).}
\label{eq:djc_edge_velocity}
\end{equation}
The amplitude localization is set by the decay of \(|\Psi|^{2}\propto e^{-2y/\xi}\),
\begin{equation}
{
\xi^{-1}
=-\Re\zeta_{\pm}
=\frac{\tilde{g}}{2}
-\frac{\tilde{g}^{2}}{1+\tilde{g}^{2}}\,\delta
+\frac{3\tilde{g}^{3}}{(1+\tilde{g}^{2})^{3}}\,\delta^{2}
+\mathcal{O}(\delta^{3}).}
\label{eq:djc_edge_xi}
\end{equation}
Define projectors onto the photon sectors,
\(\Pi_{n}=|n\rangle\langle n|\otimes\mathbb{I}_{2\times2}\) and
\(\Pi_{n+1}=|n+1\rangle\langle n+1|\otimes\mathbb{I}_{2\times2}\).
For the left-edge state \(\Psi_{L}(y)=\sum_{s=\pm}c_{s}e^{\zeta_sy}\phi_{s}\) with \(c_{+}=1,\;c_{-}=-1\), the integrated weight in photon sector \(m=n,n+1\) is
\begin{equation}
{
I_{m}= -\sum_{s,t=\pm}
\frac{c_{s}^{*}c_{t}\,\phi_{s}^{\dagger}\Pi_{m}\phi_{t}}
{\zeta_{s}^{*}+\zeta_{t}}.}
\label{eq:djc_edge_sector_weight}
\end{equation}
Normalizing by \(p_{m}=I_{m}/(I_{n}+I_{n+1})\) gives the conditional probabilities; the mean photon number is
\begin{equation}
{ \langle a^{\dagger}a\rangle = n\,p_{n}+(n+1)\,p_{n+1}.}
\label{eq:djc_edge_photon_number}
\end{equation}
The local conditional sector weights are
\begin{equation}
p_m(y)=\frac{\Psi_L^\dagger(y)\Pi_m\Psi_L(y)}{\Psi_L^\dagger(y)\Psi_L(y)},
\label{eq:djc_edge_local_sector_weight}
\end{equation}
and, unlike the integrated probabilities, they vary along the edge profile.
At the left boundary, the edge condition gives \(\Psi_L(0)\propto(0,q_n,0,q_{n+1})^{\mathsf T}\), with \(q_n=\tilde g(\zeta_+-\zeta_-)\) and \(q_{n+1}=2S\). On the physical left-edge branch satisfying $F_L(E,k_x)=0$, the local upper-sector probability is therefore
{
\begin{equation}
p_{n+1}(0;E,k_x,\tilde g)=
\frac{4|S|^2}
{\tilde g^2|\zeta_+-\zeta_-|^2+4|S|^2}
=\frac12+E.
\label{eq:djc_edge_boundary_photon_probability}
\end{equation}
}
Along the left edge branch, with \(\delta=k_x+\tilde g/2\), this becomes
{
\begin{equation}
p_{n+1}(0)=\frac12
-\frac{\tilde g}{1+\tilde g^2}\,\delta
+\frac{\tilde g^2(2-\tilde g^2)}{(1+\tilde g^2)^3}\,\delta^2
+\mathcal O(\delta^3).
\label{eq:djc_edge_boundary_photon_expansion}
\end{equation}
}
At the node the roots are
\begin{equation}
{\zeta_{\pm}=-(\tilde{g}\pm i)/2,}
\label{eq:djc_edge_crossing_zeta}
\end{equation}
and the boundary spinor evaluates to
\begin{equation}
{\Psi_{L}(0)\propto(0,\,-i\tilde{g},\,0,\,i\tilde{g})^{\mathsf{T}}.}
\label{eq:djc_edge_crossing_spinor}
\end{equation}
The integrated weights become \(p_{n}=p_{n+1}=1/2\), so the vacuum and one-photon occupations each equal \(1/2\).
The square-root equation \(F_{L}=0\) has the following quartic algebraic consequence after eliminating the radicals:
\begin{equation}
{
0=4E^{4}-4(1+\tilde{g}^{2})E^{3}
+(1+\tilde{g}^{4}+4\tilde{g}^{2}k_x^{2})E^{2}
+\tilde{g}^{2}(1+\tilde{g}^{2})E
+\frac{\tilde{g}^{4}}{4}-\tilde{g}^{2}k_x^{2}.}
\label{eq:djc_edge_left_polynomial}
\end{equation}
Eliminating the square roots can introduce more candidates, so every physical edge energy must also satisfy the unsquared equation \(F_L=0\).  The right-edge polynomial follows from the symmetry \(E\to -E\).
Restoring physical units gives the valley-relative crossing momentum $(k_x-K_x)_0=-g_D/(2v_F)$, the edge-speed magnitude $|v_{\mathrm{edge}}|=v_F\tilde g/(1+\tilde g^2)$, and the inverse amplitude localization length $g_D/(2v_F)$ at the node. The TM length must additionally be converted from transfer cells as described after Eq.~\eqref{eq:tm_xi}.

{\subsection{VI. Tensor network methods}
\label{sec:supp_dmrg_methods_fig2}
Here we introduce the tensor network methods we employed to compute the observables in
Fig.~\ref{fig:fig2}. The DMRG calculations use the Hamiltonian on the honeycomb lattice, Eqs.~(1-2) in the main text, i.e.,~the Coulomb electron-electron interaction is not included. The photon Hilbert space is
truncated to occupations $n=0,\ldots,n_{\rm ph}$. Hence, the interacting many-body problem arises from the single cavity mode that virtually couples to all electrons, and gives rise to a special tensor network.} The light-matter coupling $g$ contains the vacuum-field amplitude of the selected cavity mode. The material region is much smaller than the cavity wavelength, so the field is approximately uniform across the material and the mode volume is independent of material size (see explanation in main text). 
\subsubsection{Matrix product state and matrix product operator representation}

The tensor network calculations use a matrix product state (MPS) with one photon site
(the photon site has local dimension $n_{\rm ph}+1$) followed by a site for each
single-particle electron state.
Each electron site can be empty or occupied, and the MPS
tracks the total electron number.  For fully periodic calculations, the
electron sites are labeled by crystal momentum. The uniform cavity coupling preserves the total occupation $n_{\mathbf k}=n_{A,\mathbf k}+n_{B,\mathbf k}$ at each momentum. We exploit the resulting occupation-sector structure while {preserving} photon-mediated correlations between the electronic degrees of freedom: the MPS is not a product of single-particle momentum states.
The Hamiltonian is written as a matrix product operator (MPO) containing the photon number operator,
all bare hopping terms $\hat c_i^\dagger \hat c_j$, and all photon-electron coupling terms
$\hat a   \hat c_i^\dagger \hat c_j$ and $\hat a^\dagger \hat c_i^\dagger \hat c_j$.  {At $g=0$, we initialize the MPS from the noninteracting Slater determinant; at nonzero coupling, we optimize independently from this determinant and from the optimized state at the previous coupling. We perform 16 initial sweeps with increasing bond dimension and a small noise term, followed by up to 48 further sweeps without noise. The truncation cutoff is $10^{-12}$ during the initial sweeps and $10^{-13}$ thereafter. The maximum permitted bond dimension is initially 32 in panel (a) and 128 in panel (b), and is doubled when necessary, up to 1024. At $gt/\omega_c=0.025$ and $0.30$, we also check the stability of the energies and photon occupations upon doubling the permitted bond dimension. The energy-convergence tolerance is $10^{-10}t$, and energies obtained from the two initial states agree within $2\times10^{-8}t$.}
{The largest calculation contains $653$ fermions on a fully periodic
$36\times36$ lattice.}  Although the MPS is a one-dimensional chain of
momentum states, the underlying physical problem is two dimensional and
fully periodic. The calculation {preserves} the total particle number and the total momentum, while limiting the cavity mode to the stated photon cutoffs (see Table~\ref{tab:dmrg_charge_gap_params}).
{Note that the MPS bond dimension measures the Schmidt rank kept in the MPS across a partition; hence, it can be substantially smaller than the total particle number; e.g. an occupation-number product state has bond dimension one at any filling.}
In Figure~\ref{fig:dmrg_ordering_graphs} we show the real-space snake path versus the the momentum-space ``star" interaction graph in (b).

{
\begin{figure}[htbp]
\centering
\includegraphics[width=0.9\textwidth]{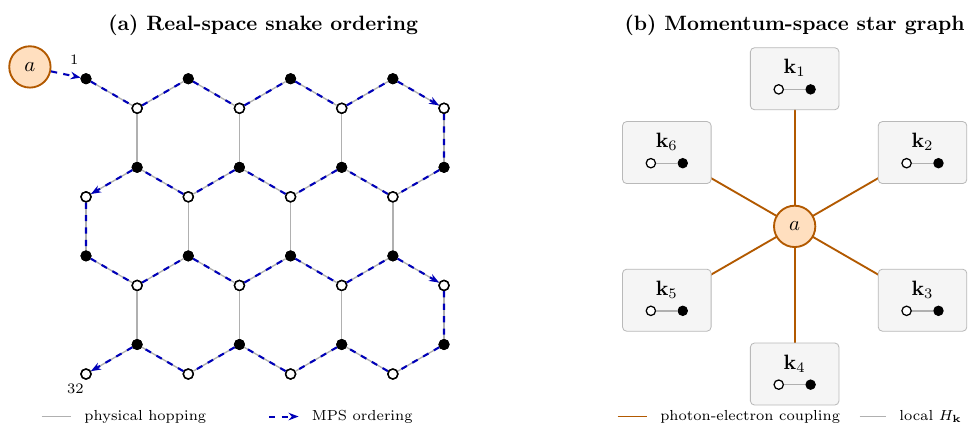}
\caption{\textbf{Real-space ordering and momentum-space interaction graph.} (a) Grey bonds show the nearest-neighbour hopping on an illustrative honeycomb patch. The dashed blue path provides the MPS ordering, starting from the photon site. Note that periodic wrap bonds are omitted here for clarity and that photon-electron coupling acts jointly on the photon and an electronic bond. (b) {Likewise, the uniform cavity mode couples to every two-sublattice momentum block (grey rectangles).} Here, grey links denote local electronic terms and orange spokes the photon-electron coupling; open and filled dots denote the $A$ and $B$ sublattices. Hence, there are no direct electronic interaction terms connecting distinct momenta. Since the shared bosonic quantum field correlates the fermionic blocks, the states do not factorize, but the total momentum is conserved. }
\label{fig:dmrg_ordering_graphs}
\end{figure}
}

\subsubsection{Numerical validation}
We ran three checks on the DMRG solver before computing the data in Fig.~\ref{fig:fig2}; the results are in Tables~\ref{tab:dmrg_exact_validation} and~\ref{tab:dmrg_photon_cutoff_validation}. First, at $g=0$ the photon decouples from the electrons, so for a fixed electron number a ground state can be chosen as a single Slater determinant with the lowest single-particle states filled. Its energy and the full exact-diagonalization (ED) energy differ by less than $4\times10^{-14}t$.
Second, the remaining rows of Table~\ref{tab:dmrg_exact_validation} compare DMRG with reference energies or charge gaps, including a noninteracting MPS check and interacting many-body ED benchmarks. The interacting systems are small enough for full many-body ED. The geometry (cylinder or torus) is listed in each row.
Third, we wrote the Hamiltonian in both real and momentum space and confirmed that ED gives identical ground-state energies for the two forms at every photon cutoff we tested. We then ran DMRG with each form. The errors in Table~\ref{tab:dmrg_photon_cutoff_validation} are measured against ED at the same finite photon cutoff. {The largest photon cutoffs tested are $n_{\rm ph}=8$ for the $4\times2$ system and $n_{\rm ph}=6$ for the $7\times2$ system.}

\begin{table}[h]

\centering
\begin{tabular}{p{0.18\textwidth}|p{0.34\textwidth}|p{0.31\textwidth}|l}
\hline\hline
test & geometry and parameters & methods compared & largest difference \\
\hline
noninteracting energy & zigzag cylinder $4\times2$, $6\times2$, $4\times4$; $g=0$ & Slater determinant and many-body ED & $<4\times10^{-14}t$ \\
interacting energy & zigzag cylinder, $2\times4$; $g=0.1$, $n_{\rm ph}=1$ & many-body ED and DMRG & $3.2\times10^{-10}t$ \\
noninteracting MPS & zigzag cylinder, $4\times8$; $g=0$, $n_{\rm ph}=0$ & Slater MPS and DMRG & $2.6\times10^{-10}t$ \\
charge gap & torus, $4\times4$; $g=0,0.05,0.10$ & many-body ED and DMRG & $2.3\times10^{-9}t$ \\
charge gap & zigzag cylinder, $8\times2$; $g=0,0.05,0.10$ & many-body ED and DMRG & $2.6\times10^{-9}t$ \\
\hline\hline
\end{tabular}
\caption{Validation of the DMRG implementation for small systems. ``Many-body ED'' means exact diagonalization of the full electron-photon Hamiltonian. The $g=0$ Slater determinant is built from the occupied eigenstates of the noninteracting electron problem.}
\label{tab:dmrg_exact_validation}
\end{table}

\begin{table}[h]

\centering
\begin{tabular}{l|c|c|c|c}
\hline\hline
system & tested $n_{\rm ph}$ & real-space DMRG error & momentum-space DMRG error & ED agreement \\
\hline
$4\times2$ & $1,\ldots,8$ & $\leq1.5\times10^{-7}t$ & $\leq1.9\times10^{-14}t$ & identical at each cutoff \\
$7\times2$ & $1,\ldots,6$ & $\leq6.5\times10^{-7}t$ & $\leq2.5\times10^{-8}t$ & identical at each cutoff \\
\hline\hline
\end{tabular}
\caption{{Comparison of ED and DMRG at the listed system sizes and photon cutoffs.}}
\label{tab:dmrg_photon_cutoff_validation}
\end{table}

\subsubsection{Panel (a) in Fig.~\ref{fig:fig2}: charge gap at half filling}

{Panel (a) uses honeycomb torus labeled by an $L\times L$ grid of lattice sites. There are $L^2$ sites, and half filling is $N_0=L^2/2$ spinless fermions.} The chosen sizes contain the Dirac momenta, so the noninteracting finite systems have zero charge gap at $g=0$.
For each coupling we compute the energy $E(N,g)$ for $N=N_0-1,N_0,N_0+1$. The particle-addition and particle-removal energies are
\begin{equation}
\mu_+(g)=E(N_0+1,g)-E(N_0,g),
\qquad
\mu_-(g)=E(N_0,g)-E(N_0-1,g).
\end{equation}
The charge gap plotted in Fig.~\ref{fig:fig2}(a) is their difference,
\begin{equation}
\Delta_c(g;L)=\mu_+(g)-\mu_-(g)
=E(N_0+1,g)+E(N_0-1,g)-2E(N_0,g),
\label{eq:dmrg_charge_gap}
\end{equation}
{$E(N,g)$ is a restricted minimum over independently optimized occupation sectors. Each DMRG run fixes every momentum occupation $n_{\mathbf k}$, while the band populations and photon dressing are free to relax. This is the construction of Eq.~\eqref{eq:e0_restricted}, with a zero-energy $\mathcal S$ and the half-filled reference $N_0=L^2/2$.}
We do not require the three states to have the same total momentum, since their momenta can differ.
{At weak coupling the gap opens quadratically in $g$, as in earlier work~\cite{Dag2024}. We use $\omega_c/t=0.1$, $n_{\rm ph}=2$, and the fixed mode volume described above. Table~\ref{tab:dmrg_charge_gap_params} lists the system sizes and DMRG parameters. Over the range $0\leq gt/\omega_c\leq0.30$ shown in Fig.~\ref{fig:fig2}(a), the finite-size curves deviate from the second-order prediction at larger coupling. They also show finite-size effects and are not ordered monotonically with $L$. The latter behavior might point to a possible gap closing in light-matter coupling, which will be systematically studied in a follow-up work.}

\begin{table}[h]

\centering
{\begin{tabular}{l|r|r|r}
\hline\hline
label & $N_0$ & $\chi_{\rm cap}$ & $\chi_{\rm actual}$ \\
\hline
$12\times12$ & 72  & 64  & 52 \\
$18\times18$ & 162 & 128 & 73 \\
$24\times24$ & 288 & 128 & 111 \\
$30\times30$ & 450 & 256 & 144 \\
$36\times36$ & 648 & 256 & 185 \\
\hline\hline
\end{tabular}}
\caption{{DMRG parameters for Fig.~\ref{fig:fig2}(a). The labels count $L^2$ lattice sites, with $N_0=L^2/2$ fermions at half filling. All sizes use $n_{\rm ph}=2$ and $\omega/t=0.1$. The two bond-dimension columns give the largest permitted value ($\chi_{\rm cap}$) and largest MPS value reached ($\chi_{\rm actual}$) among the states used for the plotted results. Additional calculations with permitted bond dimensions of 128, 256, 256, 512, and 512, respectively, were used to check convergence.}}
\label{tab:dmrg_charge_gap_params}
\end{table}

\subsubsection{Panel (b) in Fig.~\ref{fig:fig2}: charge gap near a one-photon resonance}

Panel (b) uses $L\times L$ honeycomb torus, i.e.,~periodic boundary conditions in both directions. {The size labels count lattice sites.} The calculations are above half filling. For each filling, we tune the cavity frequency at $g=0$ so that $\epsilon_-(\mathbf k_F)+\omega_c=\epsilon_+(\mathbf k_F)$, where $\epsilon_\pm$ are the noninteracting band energies at the Fermi momenta $\mathbf k_F$. {This puts the system at the noninteracting one-photon resonance; DMRG then gives the charge gap that the light--matter coupling opens at this filling.} Parameters are listed in Table~\ref{tab:dmrg_exchange_gap_params}.

\begin{table}[h]

\centering
{\begin{tabular}{l|r|r|r}
\hline\hline
label & $N_0$ & $\omega/t$ & $\chi_{\rm actual}$ \\
\hline
$12\times12$ & 76  & 1.03528 & 149 \\
$18\times18$ & 166 & 0.69459 & 169 \\
$24\times24$ & 292 & 0.52210 & 191 \\
$30\times30$ & 454 & 0.41811 & 211 \\
$36\times36$ & 652 & 0.34862 & 228 \\
\hline\hline
\end{tabular}}
\caption{{DMRG parameters for Fig.~\ref{fig:fig2}(b). The labels count $L^2$ lattice sites, and $N_0$ is the reference particle number. All sizes use $n_{\rm ph}=1$. The largest permitted bond dimension for the states used in the plotted results is $\chi_{\rm cap}=256$. The column $\chi_{\rm actual}$ gives the largest MPS bond dimension reached. Additional calculations allowing a bond dimension of 512 were used to check convergence at each size.}}
\label{tab:dmrg_exchange_gap_params}
\end{table}

{In the following, we explain how the cavity frequency should be tuned by utilizing graphene tight-binding dispersion.  
Let $\mathcal S$ denote four selected single-particle states that are numerically degenerate at $g=0$ (to within $10^{-10}t$). The clusters contain $L$ and $L/2$ primitive cells along $\bm a_1$ and $\bm a_2$, respectively. Periodic boundary conditions give
\begin{equation}
\begin{aligned}
\bm k&=\frac{u}{L}\bm b_1+\frac{2v}{L}\bm b_2,\\
u&=0,\ldots,L-1,\qquad v=0,\ldots,L/2-1,
\end{aligned}
\end{equation}
where $\bm a_i\cdot\bm b_j=2\pi\delta_{ij}$. The bare dispersion (of graphene) is
\begin{equation}
\epsilon_\pm(\bm k)=\pm t\left|1+e^{i\bm k\cdot\bm a_1}+e^{i\bm k\cdot\bm a_2}\right|.
\end{equation}
All five values of $L$ chosen in simulations are divisible by six, so the ordinary graphene Dirac points at $g=0$ formed between two photon copies} have indices $(u,v)=(L/3,L/3)$ and $(2L/3,L/6)$. Note that these are the zero-energy nodes of the electronic dispersion, not crossings between photon replicas. The four selected positive-energy states in $\mathcal S$ have indices $(L/3\pm1,L/3)$ and $(2L/3\pm1,L/6)$. The one-photon resonance is selected at these nearby momenta by tuning $\omega_c$. Substitution, with the overall phase dropping out of the absolute value, gives
\begin{equation}
\begin{aligned}
\frac{\epsilon_{\mathcal S}}{t}&=\left|e^{\pm2\pi i/L}-1\right|=2\sin(\pi/L),\\
\omega_c&=\epsilon_+(\bm k_F)-\epsilon_-(\bm k_F)=2\epsilon_{\mathcal S},
\qquad \frac{\omega_c}{t}=4\sin(\pi/L).
\end{aligned}
\end{equation}
Hence $\omega_c$ can be easily set by the system size.
{Let $\mathcal B$ denote the $N_0-2$ occupied single-particle states below the four states in $\mathcal S$.} A many-body reference configuration is specified by $\mathcal B$ together with a subset $\mathcal R\subset\mathcal S$ of occupied states from $\mathcal S$.
To find the charge gap above half filling, we optimize three groups of configurations separately: (i) all four configurations with one electron in $\mathcal S$, (ii) all six with two electrons in $\mathcal S$, and (iii) all four with three electrons in $\mathcal S$. Each configuration is a separate DMRG run with the full electron-photon MPO, restricted to its occupation sector. Minimizing over these runs independently at each particle number gives the restricted energy
\begin{equation}
E(N,g)=\min_{\substack{\mathcal R\subset\mathcal S\\|\mathcal R|=N-|\mathcal B|}}
E_{\rm DMRG}(\mathcal B\cup\mathcal R;g),
\qquad N=N_0-1,N_0,N_0+1,
\label{eq:e0_restricted}
\end{equation}
where $N_0=|\mathcal B|+2$ is the reference electron number. The minimum is chosen independently for addition, reference, and removal, and the total momenta of the three states need not coincide (in practice they do not). The addition and removal energies are $\mu_{\rm add}=E(N_0+1,g)-E(N_0,g)$ and $\mu_{\rm rem}=E(N_0,g)-E(N_0-1,g)$, and the charge gap in panel (b) is their difference, as in panel (a), albeit at a photon-exchange avoided crossing:
\begin{equation}
\Delta_c(g)=E(N_0+1,g)+E(N_0-1,g)-2E(N_0,g).\notag
\end{equation}
The uniform coupling conserves $n_{\mathbf k}=n_{A,\mathbf k}+n_{B,\mathbf k}$ at each momentum, but not the individual band populations, because the cavity interaction mixes the bands.
The full MPO allows interband transitions and photon dressing in singly occupied background momenta, so $\mathcal B$ labels the state instead of a frozen Fermi sea. At $g=0$ the common background energy cancels and $\Delta_c(0)=0$.

\subsubsection{Schrieffer-Wolff theory}
{We use Schrieffer--Wolff (SW) perturbation theory~\cite{SchriefferWolff1966} to obtain the leading second-order approximation to the charge gaps and compare with the DMRG result. For this, write $H=H_0+g(V_++V_-)$, where $H_0$ contains the uncoupled electron and photon energies and $V_\pm$ include the hopping scale $t$. The operator $V_+$ contains the photon-creation terms $\hat a^\dagger c_i^\dagger c_j$, and $V_-=V_+^\dagger$ contains the photon-annihilation terms.

We start from zero-photon Slater determinants of electrons, that are degenerate at $g=0$, as the problem setting is non-interacting without a cavity coupling. A Slater determinant is a many-electron state (in the first quantization), in which each noninteracting single-particle state is either filled or empty. This is particularly a convenient picture for us, again due to the non-interacting nature of the problem without a cavity coupling, and we have access to single-particle electronic states in graphene.

Hence, in Fig.~4(a) at $g=0$, there are four degenerate zero-energy states, because two Dirac points $\mathbf K,\mathbf K'$ each has a pair of degenerate bands. The determinants have all negative-energy states filled and one, two, or three of the four zero-energy states occupied, for $N=N_0-1,N_0,N_0+1$ respectively.

For a given particle number $N$ and total momentum $K$, let $P_{N,K}$ project onto the subspace spanned by these determinants. Let $Q_{N,K}=\sum_m\ket{m;1}\bra{m;1}$ project onto the one-photon determinants that $V_+$ can reach from $P_{N,K}$, including particle-hole excitations of the occupied background (the Fermi sea). Note that this is not the full complement $Q_{N,K}\neq 1-P_{N,K}$, but only the one-photon states contribute at second order due to the nature of light-matter coupling (1 photon processes).

Let $E_N^{(0)}$ be the common unperturbed energy of these determinants. The second-order effective Hamiltonian within $P_{N,K}$ is
\begin{equation}
\left[H^{(2)}_{N,K}\right]_{\alpha\beta}=
\sum_{m\in Q_{N,K}}
\frac{\bra{\alpha}V_-\ket{m;1}\bra{m;1}V_+\ket{\beta}}
{E_N^{(0)}-E_m^{(0)}},
\label{eq:sw_effective_matrix}
\end{equation}
where $\ket{\alpha}$ and $\ket{\beta}$ are zero-photon determinants in the subspace of $P_{N,K}$ and $E_m^{(0)}=E_{\rm el}^{(m)}+\omega_c$ includes the photon energy of an intermediate state. If $\lambda_{N,K}$ is the lowest eigenvalue of $H^{(2)}_{N,K}$, the lowest energy in this block is
\begin{equation}
E^{\rm SW}_{N,K}=E_N^{(0)}+g^2\lambda_{N,K}+O(g^4).\label{eq:fullSWenergy}
\end{equation}
We take $\lambda_N^{\min}=\min_K\lambda_{N,K}$ separately for each particle number, over all allowed total momenta and all allowed sets of electron numbers $\{n_{\mathbf k}\}$, where $n_{\mathbf k}=0,1,2$ is the number of electrons at momentum $\mathbf k$. The uniform cavity mode transfers no momentum, so an electron keeps its $\mathbf k$ when it interacts with a photon. Because $H_0$ and $V_\pm$ therefore preserve every $n_{\mathbf k}$, $H^{(2)}_{N,K}$ is block diagonal in these occupations. For determinants with definite momentum occupations,
\[
\bigl(n_{\mathbf k}^{\alpha}-n_{\mathbf k}^{\beta}\bigr)
\left[H^{(2)}_{N,K}\right]_{\alpha\beta}=0
\qquad\text{for every }\mathbf k.
\]
In both panels of Fig.~4 in the Letter, the charge gap trivially vanishes at $g=0$, $E_{N_0+1}^{(0)}+E_{N_0-1}^{(0)}-2E_{N_0}^{(0)}=0$ due to degeneracy discussed above. This is also clear from single-electron band structures. Eqs.~\eqref{eq:dmrg_charge_gap} and~\eqref{eq:fullSWenergy} then give
\begin{equation}
\Delta_c^{\rm SW}(g)=g^2\bigl(\lambda_{N_0+1}^{\min}+\lambda_{N_0-1}^{\min}-2\lambda_{N_0}^{\min}\bigr)+O(g^4).
\label{eq:sw_exchange_gap}
\end{equation}
For panel (a), the charge gap follows from counting virtual processes at the two Dirac points. At each Dirac point, an electron in one sublattice state can emit a photon and move to the empty other sublattice state, which has the same bare energy. The intermediate state therefore costs only $\omega_c$, and this process lowers the energy by $a=g_D^2/\omega_c=9t^2g^2/(2\omega_c)$, using $g_D=3tg/\sqrt2$. The process is blocked if the other sublattice state is also occupied. In the lowest configurations, one such process is available for $N_0-1$ and $N_0+1$ electrons, and two (one per Dirac point) for $N_0$ electrons. The filled negative-energy states shift all three energies equally. Hence $\Delta_c^{\rm SW}=(-a)+(-a)-2(-2a)=2a$: adding or removing an electron each costs $a$. This gives the black dashed curve, $\Delta_c^{\rm SW}/\omega_c=9(gt/\omega_c)^2+O((gt/\omega_c)^4)$, with the same coefficient for all five sizes.}

In the perturbative regime, it is straightforward to understand why SW transformation can capture the charge gap at the $\mathbf{K}$ point (half-filling), as the cavity field merely acts like a time-reversal symmetry breaking background for the electronic system. When the cavity field is a mere background, projecting to vacuum, as seen in Eq.~\eqref{eq:sw_effective_matrix} provides a correct treatment for the many-body problem, and we do find matching results with DMRG in weak light-matter interaction strength. The same is not immediately obvious at the points where different photon copies hybridize, giving rise to the photon-exchange gaps in which the bright edge modes reside and we study in detail in this work. This is because the states where a photon-exchange gap opens at $g>0$ are hybrid, an entangled electron-photon state, and hence not in vacuum. However the states in the close vicinity exhibit product-like state character, either in vacuum or in 1-photon sector, similar to the states at the graphene valleys. Concretely, if we project to vacuum, we obtain an electronic (charge) gap in the region of momentum space where photon-exchange gap resides. This is the gap which we can apply SW transformation to, and in fact capture in the DMRG calculations, seen in Fig.~4(b).

For panel (b), the reference determinants contain again no photon, but photons still enter through the one-photon intermediate states in Eq.~\eqref{eq:sw_effective_matrix}. At the chosen filling, the lower-band state at each resonant momentum is occupied, so Pauli exclusion blocks the resonant photon emission from $\mathcal S$ and the energy denominators stay finite. The remaining virtual processes shift the three charge sectors differently, which gives a charge gap at the order $g^2$. Let us emphasize that this scaling is different from the photon-exchange gap for a single electron on an empty lattice, i.e.,~$\Delta_{1\rm p} \sim gt$ (linear in light-matter coupling strength). This is understandable, because the nature of these gaps, one an electronic charge gap and the other a hybrid photon-exchange gap, is different.  Then, we can apply the same second-order perturbation theory to the positive-energy states $\mathcal S$ and background $\mathcal B$. The projector onto the corresponding zero-photon subspace is
\begin{equation}
P_{N,K}=
\sum_{\substack{\mathcal R\subset\mathcal S,\
|\mathcal R|=N-|\mathcal B|\\
K(\mathcal B\cup\mathcal R)=K}}
\ket{\mathcal B\cup\mathcal R;0}\bra{\mathcal B\cup\mathcal R;0}.
\label{eq:sw_projector}
\end{equation}

The effective Hamiltonian and charge gap are given by Eqs.~\eqref{eq:sw_effective_matrix} and~\eqref{eq:sw_exchange_gap}.
Different subsets $\mathcal R\subset\mathcal S$ put different numbers of electrons at the individual momenta, so the off-diagonal matrix elements between them vanish even when they have the same total momentum $K$. All four zero-energy Dirac-point states are filled in $\mathcal B$, so no further degeneracy arises for a given occupation of $\mathcal S$. We therefore compute the second-order shifts separately for each configuration and then minimize separately at each of the three particle numbers. Photon-mediated correlations at fixed $\{n_{\mathbf k}\}$ are included.

{This is the leading contribution from virtual one-photon processes. Table~\ref{tab:sw_sector_shifts} lists the minimized shifts for all five sizes and the combination $C_{\rm SW}=\lambda_{N_0+1}^{\min}+\lambda_{N_0-1}^{\min}-2\lambda_{N_0}^{\min}$, in units of $t$.}
\begin{table}[h]

\centering
{\begin{tabular}{l|r|r|r|r}
\hline\hline
system & $\lambda_{N_0-1}^{\min}/t$ & $\lambda_{N_0}^{\min}/t$ & $\lambda_{N_0+1}^{\min}/t$ & $C_{\rm SW}/t$ \\
\hline
$12\times12$ & $-14.42033752$ & $-16.87818486$ & $-13.39837572$ & $5.93765648$ \\
$18\times18$ & $-38.76212335$ & $-42.63761067$ & $-37.18588120$ & $9.32721679$ \\
$24\times24$ & $-75.90660297$ & $-81.16292663$ & $-73.78449457$ & $12.63475573$ \\
$30\times30$ & $-126.20367199$ & $-132.82578025$ & $-123.53914941$ & $15.90873910$ \\
$36\times36$ & $-189.86389066$ & $-197.84418954$ & $-186.65869774$ & $19.16579068$ \\
\hline\hline
\end{tabular}}
\caption{{Lowest second-order energy shifts for $N_0-1$, $N_0$, and $N_0+1$ electrons, in units of $t$. The interaction operators include $t$, and $g$ is dimensionless, so the shifts and $C_{\rm SW}$ have units of energy.}}
\label{tab:sw_sector_shifts}
\end{table}
With these shifts, Eq.~\eqref{eq:sw_exchange_gap} gives
\begin{equation}
\frac{\Delta_c^{\rm SW}}{\omega_c}=A_L(gt/\omega_c)^2+O((gt/\omega_c)^4),
\qquad
A_L=\frac{C_{\rm SW}\omega_c}{t^2}.
\end{equation}
The coefficients are $A_L=(6.14711432,\,6.47861679,\,6.59666622,\,6.65166420,\,6.68163489)$ for $L=(12,18,24,30,36)$.
{Table~\ref{tab:sw_exchange_validation} compares this second-order prediction, which has no fit parameters, with DMRG at $gt/\omega_c=0.025$ and $0.050$. The two agree at weak coupling, which indicates that virtual one-photon processes give the leading quadratic term. Deviations at larger coupling are consistent with corrections beyond second-order SW theory.}
\begin{table}[h]

\centering
{\begin{tabular}{c|c|c|c|c}
\hline\hline
system & $gt/\omega_c$ & $\Delta_c^{\rm DMRG}/\omega_c$ & $\Delta_c^{\rm SW}/\omega_c$ & $(\Delta_c^{\rm DMRG}-\Delta_c^{\rm SW})/\omega_c$ \\
\hline
$12\times12$ & 0.025 & 0.003843233 & 0.003841946 & $+1.287\times10^{-6}$ \\
$12\times12$ & 0.050 & 0.015381669 & 0.015367786 & $+1.388\times10^{-5}$ \\
$18\times18$ & 0.025 & 0.004052416 & 0.004049135 & $+3.280\times10^{-6}$ \\
$18\times18$ & 0.050 & 0.016239373 & 0.016196542 & $+4.283\times10^{-5}$ \\
$24\times24$ & 0.025 & 0.004127314 & 0.004122916 & $+4.397\times10^{-6}$ \\
$24\times24$ & 0.050 & 0.016550641 & 0.016491666 & $+5.897\times10^{-5}$ \\
$30\times30$ & 0.025 & 0.004162350 & 0.004157290 & $+5.060\times10^{-6}$ \\
$30\times30$ & 0.050 & 0.016697652 & 0.016629161 & $+6.849\times10^{-5}$ \\
$36\times36$ & 0.025 & 0.004181503 & 0.004176022 & $+5.481\times10^{-6}$ \\
$36\times36$ & 0.050 & 0.016778594 & 0.016704087 & $+7.451\times10^{-5}$ \\
\hline\hline
\end{tabular}}
\caption{{DMRG compared with the second-order SW prediction without fit parameters. Both gaps are normalized by $\omega_c$, and the coupling is $gt/\omega_c$, as in the figure.}}
\label{tab:sw_exchange_validation}
\end{table}

\end{widetext}

\end{document}